\documentclass[12pt]{article}
\usepackage{amscd,amssymb,amsmath,latexsym,enumerate}
\usepackage[mathscr]{euscript}
\usepackage{mathrsfs}
\usepackage{epsfig}
\usepackage{fancybox}
\usepackage{verbatim}
\usepackage{tikz}
\usepackage{tikz-cd}
\usetikzlibrary{arrows}

\usepackage{todonotes}
\usepackage{multicol}
\usepackage{graphicx}

\usepackage{color}

\usepackage{xcolor}
\definecolor{MyBlue}{cmyk}{1,0.13,0,0.63}
\definecolor{MyGreen}{cmyk}{0.91,0,0.88,0.52}
\newcommand{\mylinkcolor}{MyBlue}
\newcommand{\mycitecolor}{MyGreen}
\newcommand{\myurlcolor}{black}

\usepackage{hyperref}
\hypersetup{%
  bookmarksnumbered=true,bookmarksopen=false,%
  plainpages=false,% necessary to prevent duplicate page identifiers
  linktocpage=true,%
  colorlinks=true,breaklinks=true,%
  linkcolor=\mylinkcolor,citecolor=\mycitecolor,urlcolor=\myurlcolor,%
  pdfpagelayout=OneColumn,%
  pageanchor=true,%
}

\title{Detecting local topology with the spectral localizer}

\author{Alexander Cerjan$^{1}$, 
Hermann Schulz-Baldes$^2$
\\
\\
{\small $^1$Center for Integrated Nanotechnologies, Sandia National Laboratories}
\\
{\small Albuquerque, New Mexico 87185, USA}
\\
{\small $^2$FAU Erlangen-N\"urnberg, Department Mathematik, Cauerstr. 11, D-91058 Erlangen, Germany}
}

\date{ }

\newtheorem{theorem}{Theorem}
\newtheorem{proposition}[theorem]{Proposition}
\newtheorem{lemma}[theorem]{Lemma}

\newtheorem{definition}[theorem]{Definition}
\newtheorem{remark}[theorem]{Remark}

\newcommand{\CM}{{\mathbb C}}
\newcommand{\NM}{{\mathbb N}}
\newcommand{\RM}{{\mathbb R}}

\newcommand{\ZM}{{\mathbb Z}}

\newcommand{\EM}{{\mathbb E}}

\newcommand{\Oo}{{\cal O}}

\newcommand{\Nn}{{\cal N}}

\newcommand{\Hh}{{\cal H}}

\newcommand{\HP}{{H'}}

\newcommand{\one}{{\bf 1}}

\newcommand{\spec}{\mbox{\rm spec}}

\newcommand{\SF}{{\rm Sf}}

\newcommand{\Ran}{{\rm Ran}}

\newcommand{\Sig}{{\rm Sig}} 
\newcommand{\LocInd}{{\rm Ch}} 
\newcommand{\diag}{{\rm diag}} 
\newcommand{\supp}{{\rm supp}} 
\newcommand{\dist}{{\rm dist}}

\begin{document}

\maketitle

%%%%%%%%%%%%%%%%%%%%%%%%%%%%%%%%%%%%%%%%%%%%%%%%%%%%
\begin{abstract}
The spectral localizer is a predictive framework for the computation of topological invariants of natural and artificial materials. Here, three crucial improvements on the criterion for the validity of the framework are reported: first, merely a properly defined local spectral gap of the Hamiltonian is required, second, only relative bounds on the Hamiltonian and its noncommutative derivative are relevant, and, third, the numerical constant in a tapering estimate is improved. These developments further stress the local nature of the spectral localizer framework, enabling more precise predictions in heterostructures, aperiodic, and disordered systems. Moreover, these results strengthen the bounds on the spectral localizer's spectral flow when crossing topological phase boundaries.

\vspace{.1cm}


%\noindent {AMS MSC2010:} 47B36, 39A21, 37C30

%\noindent {\bf Acknowledgements:} This work was supported by the DFG grant SCHU 1358/6-2. 

\end{abstract}
%%%%%%%%%%%%%%%%%%

%\newpage

%\tableofcontents

%%%%%%%%%%%%%%%%%%%%%%%%%%%%%%%%%%%%%%%%%%%%%
\section{Introduction}

Topological materials are often described by gapped periodic Hamiltonians. Extensions to disordered or, more generally, space-inhomogeneous systems are then possible using the non-commutative Brillouin zone \cite{PS}. In this non-commutative Brillouin zone framework, index theorems connect the topological invariants of crystalline system to indices of suitably constructed Fredholm operators describing the associated aperiodic system. Likely the most efficient tool for the numerical computation of these invariants in such aperiodic systems is the spectral localizer \cite{LS1,LS2}. The aim of this work is to study heterostructures that have local topological invariants and possibly topological phase boundaries between them. Such systems cannot be described by homogenous operators for which the techniques of \cite{PS} apply. A typical example to have in mind are interfaces between two insulators with different material topology. One way to approach the study of heterostructure Hamiltonians is via the integral kernel of the index approach, also called the local marker {\cite{BRe}. But} here we rather use the spectral localizer to define the local topological invariants. This spectral localizer-based approach has the great advantage of being readily implemented numerically, while local markers and indices of Fredholm operators are harder to compute. The spectral localizer framework has found a wide range of applications in the physics community \cite{Fulga16, CL0, Franca,CL24}, especially because it permits efficient numerical implementations that can be applied directly to experimentally realizable systems without requiring a low-energy description \cite{Cheng, Dixon, Wong, Ochkan, Spataru}. 

\vspace{.2cm}

This work adds three essential improvements on prior works about the spectral localizer framework. First, it is shown that the existence of local topological invariants is guaranteed if the bounded Hamiltonian $H$ has merely a local spectral gap $g_\rho>0$ at energy $E$, which by definition means that the restriction of $(H-E\,\one)^2$ to a finite volume parameterized by the length $\rho$ is strictly positive. This new notion of a local spectral gap is of considerable interest on its own and is presented and discussed in Section~\ref{sec-LocalGapHam}. Second, a new quantitative criterion on the length $\rho$ and on the tuning parameter $\kappa$ of the spectral localizer is presented, which then assures the existence of a gap of the spectral localizer. This criterion involves relative operator bounds on the Hamiltonian and its noncommutative derivative w.r.t.\ the position operator, and allows one to prove in a quantitative manner the locality properties of the spectral localizer. The third improvements concerns so-called tapering estimates which lead to better constants in the criterium for topological protection.

\vspace{.2cm}

While the technical description is deferred to later, let us in this introductory section be somewhat more explicit for the case of {tight-binding models in} dimension $d=2$. Then $H$ acts on $\ell^2(\ZM^2,\CM^L)$ where $L$ is the number of internal degrees of freedom, and on this Hilbert space there are given the two components $X_1,X_2$ of the position operator. The (even) spectral localizer is then defined as the operator
\begin{equation}
\label{eq-SpecLoc2d}
L_{\kappa}
\;=\;
\begin{pmatrix}
-(H-E\,\one) & \kappa(X_1-\imath X_2) \\
 \kappa(X_1+\imath X_2) & H-E\,\one
\end{pmatrix}
\;,
\end{equation}
and its finite volume restriction is a finite-dimensional selfadjoint matrix denoted by $L_{\kappa,\rho}$. Provided it has a spectral gap at $0$, the local topological index (or local Chern marker) is then defined as the half-signature of $L_{\kappa,\rho}$ by
$$
\LocInd_{\kappa,\rho}
\;=\;
\frac{1}{2}\,\Sig(L_{\kappa,\rho})
\;.
$$
The terminology {\it local Chern marker} is justified by the main result of Ref.~\cite{LS2}, which shows that when $H$ is a gapped homogeneous Hamiltonian for which the Chern number is a well-defined integer \cite{PS}, this bulk Chern number is precisely equal to $\LocInd_{\kappa,\rho}$ (see also Section~10 of \cite{DSW} for a simplified proof based on spectral flow). As such, guaranteeing that the spectral gap of $L_{\kappa,\rho}$ is open is of crucial importance to classifying a system's local topology, and the new quantitative criterion, ensuring that the gap of $L_{\kappa,\rho}$ is at least half the local gap $g_\rho$, is
\begin{equation}
\label{eq-Criterion2d}
\frac{2\,g_\rho}{\rho}\;<\;\kappa\;<\; 
\frac{g_\rho^3}{\frac{5}{3}\big(2\,\|H(\imath \,\one +\frac{1}{\rho} |X|)^{-1}\|+g_\rho\big)\,\|[ X_1+\imath X_2,H](\imath \,\one +\frac{1}{\rho}  |X|)^{-1}\|}
\;,
\end{equation}
where $|X|=(X_1^2+X_2^2)^{\frac{1}{2}}$. The detailed statement for arbitrary even-dimensional systems is given in Theorem~\ref{theo-localGap} in Section~\ref{sec-SpecLocGap}, which also contains a thorough discussion of this main result of the paper. As already stressed above, the new development here compared to \cite{LS2,DSW} is that only the local gap $g_\rho$ enters (and {\em not} the global gap) and that the norms of $H$ and its commutator with position are replaced by the relative operator norm bounds, namely there appears the resolvent of $|X|$ in the above criterion. Furthermore, the constants in the inequality are considerably improved, see the discussion in Section~\ref{sec-SpecLocGap}. As is explained in Section~\ref{sec-StabGap}, these improvements considerably strengthen the locality properties of the local topological indices. In particular, a perturbation of $H$ by possibly large terms far away from the origin is damped by these resolvents and hence barely influences the criterion. Further stability results of this work concern the spectral flow of the spectral localizer if it is moved over a topological phase boundary, see Sections~\ref{sec-SFstab} and \ref{sec-SFstab2}. All of the above results are illustrated by numerical examples in Section~\ref{sec-Numerics}.

\vspace{.2cm}

This work considers the even spectral localizer, which enables the computation of (higher) Chern numbers and hence, the strong invariants in even dimensional physical systems. However, the arguments transpose directly to the odd spectral localizer (described in \cite{LS1,DSW}). Moreover, the new criterion \eqref{eq-Criterion2d} also applies to all other situations in which the spectral localizer can be used to compute index invariants. For example, if the Hamiltonian has real symmetries (such as time-reversal and particle-hole invariance), or a combination of crystalline and local symmetries that yield fragile topology \cite{Ho}, there may be $\ZM_2$-valued torsion invariants which can then be computed as sign of the Pfaffian of real skew-adjoint version of the spectral localizer \cite{DoSB,Lee25}, and based on the results of this paper this connection now holds under the weaker condition \eqref{eq-Criterion2d}. The same can be said about integer-valued invariants (such as Spin-Chern numbers) in presence of approximate symmetries \cite{DoSB2}, as well as weak invariants computed as semi-finite index pairings \cite{SS}. More generally and from a mathematical perspective, the weaker criterion (supposing the stronger hypothesis invoking a global gap of $H$) applies to any index pairing of a $K$-theory class with an unbounded Dirac operator specifying a $K$-homology class and then allows to compute the pairing via the spectral localizer under criteria which only invoke relative operator bounds \cite{DSW}.

%%%%%%%%%%%%%%%%%%%%%%%%%%%%%%%%%%%%%%%%%%%%%
\section{Local spectral gaps of the Hamiltonian}
\label{sec-LocalGapHam}

Let us consider the following concrete set-up. The Hilbert space is chosen to be $\ell^2(\ZM^d,\CM^L)$ over a $d$-dimensional tight-binding lattice with $L$ orbitals per site. On this Hilbert space act the selfadjoint position operators $X_1,\ldots,X_d$, defined by $X_j|n,l\rangle =n_j|n,l\rangle$ where $n=(n_1,\ldots,n_d)\in\ZM^d$ and $l\in\{1,\ldots,L\}$, and $|n,l\rangle$ denotes the state over site $n$ in the $l$-th orbital. Given an irreducible selfadjoint representation $\gamma_1,\ldots,\gamma_d$ of the Clifford algebra $\CM^{d'}$ with $d'=2^{\lfloor\frac{d}{2}\rfloor}$, the (dual) Dirac operator centered at $x=(x_1,\ldots,x_d)\in\RM^d$ is then defined {as an operator on $\ell^2(\ZM^d,\CM^L)\otimes\CM^{d'}\cong \ell^2(\ZM^d,\CM^{Ld'})$} by 
$$
D(x)
\;=\;
\sum_{j=1}^d{(X_j-x_j)\otimes \gamma_j}
\;.
$$ 
Let us also set $D=D(0)$ and note that the domain of $D(x)$ is independent of $x$. Now the Hamiltonian $H=H^*$ is a bounded tight-binding model on $\ell^2(\ZM^d,\CM^L)$. {Its extension to $\ell^2(\ZM^d,\CM^{Ld'})$ is denoted by $\HP=H\otimes\one_{d'}$}. The following weak locality assumption on $H$ will be assumed throughout:

%%%%%%%%%%%%%%%%%%%%
\begin{definition}
The Hamiltonian is called weakly local if $\HP$ leaves the domain of $D(x)$ invariant and $[D(x),\HP]$ extends to a bounded operator. It is called uniformly weakly local if the norms  $\|[D(x),\HP]\|$ are uniformly bounded in $x$.
\end{definition}
%%%%%%%%%%%%%%%%%%%%

In terms of non-commutative geometry, this means that $\HP$ is a differential element w.r.t.\ $D$. Let us give an easy criterion for weak locality. If the off-diagonal matrix elements of $H$ satisfy $\|\langle m|H|n\rangle\|\leq C(1+|n-m|)^{-(1+\delta)}$ for some $\delta>0$ and uniform constant $C$, then $H$ is uniformly weakly local. {Of course, the latter condition holds for Hamiltonians of so-called finite range which by definition satisfy $\langle m|H|n\rangle=0$ for $|n-m|>r$ for some finite $r$. This holds for most tight-binding models studied in the physics literature.}

\vspace{.2cm}

Next let us introduce the following objects: For any $\rho>0$ and $x\in\RM^d$, let $B_\rho(x)$ be either the Euclidean ball $\{y\in\RM^d\,:\,\|x-y\|_2<\rho\}$ of radius $\rho$  centered at $x$, or the square box $\{y\in\RM^d\,:\,\|x-y\|_\infty<\frac{\rho}{2}\}$ of side length $\rho$ around $x$. Let then $\pi_\rho(x):\ell^2(\ZM^d ,\CM^L)\to \ell^2(\ZM^d \cap B_\rho(x),\CM^L)$ denote the natural surjective partial isometry. Hence its adjoint $\pi_\rho(x)^*$ is the injective partial isometry embedding the finite dimensional Hilbert space over the ball or box into $\ell^2(\ZM^d,\CM^L)$. Then $\one_\rho(x)=\pi_\rho(x)^*\pi_\rho(x)$ is a multiplication operator on $\ell^2(\ZM^d,\CM^L)$, namely by the indicator function on the ball or box. Let us set $A_{\rho}(x)=\pi_\rho(x)A\pi_\rho(x)^*$ for any operator $A$ on $\ell^2(\ZM^d,\CM^L)$. Hence $A_\rho(x)$ is the Dirichlet restriction of $A$ to $\ZM^d\cap B_\rho(x)$.

%%%%%%%%%%%%%%%%%%%%
\begin{definition}
\label{def-LocalGapDirichlet}
A Hamiltonian is said to have a $\rho$-local gap at $x\in\RM^d$ if and only if there exists a constant $g>0$ such that
\begin{equation}
\label{eq-LocalGapDirichlet}
(H^2)_\rho(x)\;\geq \;g^2\,\one_\rho(x)
\;.
\end{equation}
The $\rho$-local gap $g_{\rho}(H,x)$ of $H$ at $x$ is defined to be the largest $g$ such that \eqref{eq-LocalGapDirichlet} holds, namely
\begin{equation}
\label{eq-LocalGapDef}
g_\rho(H,x)\;=\; 
\inf \spec\big((H^2)_\rho(x)\big)
\;.
\end{equation}
\end{definition}
%%%%%%%%%%%%%%%%%%%%

The main idea behind Definition~\ref{def-LocalGapDirichlet} is to work with the restriction of the square $H^2$ of $H$, which is {\bf not} equal to the square of the restriction. In fact, $(H^2)_\rho$ will not see any surface or interface modes if $B_\rho(x)$ does not intersect a surface or interface (a statement that does {\bf not} hold for $(H_\rho)^2$). On a computational level, it is very easy to compute $g_\rho(H,x)$ based on \eqref{eq-LocalGapDef}. Let us state a few very elementary facts about the $\rho$-local gap, {which are numerically illustrated in Fig.~\ref{fig:gapDemo}.}

%%%%%%%%%%%%%%%%%%%%%%%%
\begin{proposition}
\label{prop-LocalGap}
Let $H$ be weakly local.
\begin{itemize}

\item[{\rm (i)}] If $H$ has a global gap $g>0$, {\it i.e.}\ $H^2\geq g^2$, then $g_{\rho}(H,x)\geq g$ for all $\rho$ and $x$, namely $H$ has a $\rho$-local gap at $x$ for all $\rho$ and $x$.  

\item[{\rm (ii)}] For all $\rho'\leq \rho$, one has $g_{\rho'}(H,x)\geq g_{\rho}(H,x)$.

\item[{\rm (iii)}] Let the support $\supp(W)=\{n\in\ZM^d\,:\,\langle m|W|n\rangle\not=0\;\mbox{ for some }m\in\ZM^d\}$ of a bounded  operator $W$ on $\ell^2(\ZM^d,\CM^L)$ satisfy  $\supp(W)\cap B_\rho(x)=\emptyset$. Then 
$$
g_{\rho}(H+W,x)
\;=\;
g_{\rho}(H,x)
\;.
$$
In particular, if $H$ has a $\rho$-local gap at $x$, then so does $H+W$. 

\end{itemize}
\end{proposition}
%%%%%%%%%%%%%%%%%%%%%%%%

\noindent {\bf Proof.} (i) and (iii) are obvious, and (ii) follows the fact that $(H^2)_{\rho'}(x)$ is a diagonal submatrix of the non-negative matrix $(H^2)_\rho(x)$.
\hfill $\Box$

\vspace{.2cm}

Proposition~\ref{prop-LocalGap}(iii) can be read as a stability result on local spectral gaps: if the perturbation $W$ is outside of $B_\rho(x)$, then local gap remains open (actually, even its value remains unchanged). Of course, one can also consider a perturbation $W=W^*$ with a support that has a non-trivial intersection with $B_\rho(x)$. Suppose that $W$ is centered near some point $y\in\ZM^d$ and is decaying away from it such that $\|W(\one+|X-y|)\|=\|(\one+|X-y|)W\|<\infty$ is roughly the same as $\|W\|$. Setting $R=(\one+|X-y|)^{-1}$, one then has the estimate
\begin{align*}
& 
\big\|\pi_\rho(x)(H+W)^2\pi_\rho(x)^*
\,-\,
\pi_\rho(x)H^2\pi_\rho(x)^*\big\|
\;=\;
\big\|\pi_\rho(x)(HW+WH+W^2)\pi_\rho(x)^*\big\|
\\
&
\;=\;
\big\|\pi_\rho(x)(HW(\one+|X-y|)R+R(\one+|X-y|)WH+W^2(\one+|X-y|)R)\pi_\rho(x)^*\big\|
\\
&
\;\leq\;
\frac{1}{\dist(y,B_\rho(x))}\,\|(\one+|X-y|)W\|\,
\big({2\,\|H\|}+\|W\|\big)
\;,
\end{align*}
so that  
\begin{equation}
\label{eq-WeylType}
g_{\rho}(H+W,x)\;\geq\; g_{\rho}(H,x)\;-\;\frac{1}{\dist(y,B_\rho(x))}\,\|(\one+|X-y|)W\|\,\big(2\|H\|+\|W\|\big)
\end{equation}
holds. Thus, for $W$ small enough or sufficiently far away from $x$, the $\rho$-local gap at $x$ remains open, even if $W$ has some support in $B_\rho(x)$.

\vspace{.2cm}

Let us next come to a partial inverse to Proposition~\ref{prop-LocalGap}(i), {under the restrictive condition of a Hamiltonian of finite range.}

%%%%%%%%%%%%%%%%%%%%%%%%
\begin{proposition}
\label{prop-LocalToGlobalGap}
Let $H$ be {of finite range}. If there is a $\rho$ sufficiently large such that $H$ has a $\rho$-local gap uniformly for all $x$, then {$H$ has a global gap}. More quantitatively, if $g_{\rho}(H)=\inf_{x\in\RM^d}g_{\rho}(H,x)>0$ {for $\rho$ sufficiently large}, then there is a constant $c_d>0$ such that $H^2\geq (c_d\,g_{\rho}(H))^2$. 
\end{proposition}
%%%%%%%%%%%%%%%%%%%%%%%%

Some technical preparations for the proof are needed. Let $F:\RM\to [0,1]$ be an even continuously differentiable satisfying $F(y)=0$ for $|y|\geq 1$ and $F(y)=1$ for $|y|\leq \frac{1}{2}$. Given $\rho>0$, furthermore set $F_\rho(y)=F(\frac{y}{\rho})$. Then the bound \eqref{eq-LocalGapDirichlet} directly implies that
\begin{equation}
\label{eq-LocalGap}
F_{\rho}(D(x))\,(\HP)^2\,F_{\rho}(D(x))\;\geq \;g_\rho(H,x)^2\,(F_{\rho}(D(x)))^2
\;.
\end{equation}
As explained below, this bound is the crucial ingredient for the proof of topological protection in \cite{LS1,LS2,DSW}. In these works it was obtained as a consequence of a global gap on $H$, but it clearly also follows from the weaker condition of having a $\rho$-local gap. The bound \eqref{eq-LocalGap} is an operator inequality between two positive operators, namely the difference between left and right is a positive operator. Furthermore, the operator $F_{\rho}(D(x))$ is defined via spectral calculus, and as $F$ is even, one has $F_{\rho}(D(x))=F_{\rho}(|D(x)|)$. Unfolding the definitions for $x=(x_1,\ldots,x_d)\in\RM^d$, one obtains 
$$
F_{\rho}(D(x))
\;=\;
F\Big(\tfrac{1}{\rho}\big(\sum_{j=1}^d(X_j-x_j)^2{\otimes \one_{d'}}\big)^{\frac{1}{2}}\Big)
\;,
$$
namely $F_{\rho}(D(x))$ is a smoothened indicator function on balls of size $\rho$.  It is shown in \cite{LS2} (see also proof of Theorem~10.3.1 in \cite{DSW}) that the function $F$ can be chosen such that one has the commutator bound
\begin{equation}
\label{eq-TaperingEstOld}
\|[F_\rho(D(x)),\HP]\|
\;\leq\;
\frac{8}{\rho}\,
\|[D(x),\HP]\|
\;.
\end{equation}
The proof of \eqref{eq-TaperingEstOld}, actually with an improved constant, is reproduced in Appendix~\ref{app-tapering}.

\vspace{.2cm}

\noindent {\bf Proof} of Proposition~\ref{prop-LocalToGlobalGap}. As already stated above, the existence of a $\rho$-local gap implies that \eqref{eq-LocalGap} holds. Let us introduce the notation {$F_{ \rho,sn}=F_{ \rho}(D(sn))$. Noting that a cube of side-length $s=\frac{\rho}{2\sqrt{d}}$ fits inside a sphere of radius $\frac{\rho}{2}$, obtains the lower  bound of
$$
\one
\;\leq\;
\sum_{n\in\ZM^d}
(F_{ \rho,sn})^2
\;\leq\;
c'_d\,\one
\;.
$$
The upper bounds with some constant $c'_d$ follows by counting the number of functions in the sum having $x$ in its support. Then }
\begin{align*}
(\HP)^2
&
\;\geq\;
\HP
\tfrac{1}{c'_d}
\Big(\sum_{n\in\ZM^d}
(F_{ \rho,sn})^2\Big)
\HP
\\
&
\;=\;
\tfrac{1}{c'_d}
\sum_{n\in\ZM^d}
\big(
F_{ \rho,sn} (\HP)^2 F_{ \rho,sn} +
[\HP,F_{ \rho,sn}] F_{ \rho,sn} \HP+
F_{ \rho,sn} \HP[ F_{ \rho,sn},\HP]
\big)
\\
&
{
\;\geq\;
\tfrac{1}{c'_d}
\sum_{n\in\ZM^d}
\big(
g_\rho(H)^2
\,(F_{ \rho,sn})^2 +
[\HP,F_{ \rho,sn}] F_{ \rho,sn} \HP+
F_{ \rho,sn} \HP[ F_{ \rho,sn},\HP]
\big)
\;,
}
\end{align*}
\noindent {where the last bound follows from \eqref{eq-LocalGap} and the definition of $g_\rho(H)$ in the statement of the proposition. Now let us use the hypothesis that $H$ has a finite range $r$. Then $F_{ \rho,sn} \HP=F_{ \rho,sn} \HP\chi_{\rho+r,sn}$ where $\chi_{\rho+r,sn}$ is the indicator function (orthogonal projection) on states lying over sites in a ball of radius $\rho+r$ centered around $sn$. Hence
$$
[\HP,F_{ \rho,sn}] F_{ \rho,sn} \HP
\;=\;
\chi_{\rho+r,sn} [\HP,F_{ \rho,sn}] F_{ \rho,sn} \HP\chi_{\rho+r,sn}
\;\leq\;
\chi_{\rho+r,sn} \|[\HP,F_{ \rho,sn}] F_{ \rho,sn} \HP\| \chi_{\rho+r,sn}
\;.
$$
Combining this with \eqref{eq-TaperingEstOld} then leads to
}
\begin{align*}
(\HP)^2
&
\;\geq\;
\tfrac{1}{c'_d}
\sum_{n\in\ZM^d}
\big(
{g_\rho(H)^2}\,F_{ \rho,sn}^2 -
2\,{\chi_{ \rho+r,sn}}\,\|\HP\|\,\tfrac{8}{\rho}\,\|[D,\HP]\|
\big)
\\
&
\;\geq\;
{\big(\tfrac{1}{c'_d}\,{g_\rho(H)^2}-2\,c''_d\|H\|\,\tfrac{8}{\rho}\,\|[D,\HP]\|\big)\one}
\;,
\end{align*}
{where the above lower bound on the sum $\sum_{n\in\ZM^d}
(F_{ \rho,sn})^2$ was used and $c''_d$ counts the number of non-vanishing $\chi_{ \rho+r,sn}$ over a each summand $n$. The last bound} implies the claim, with a concrete (albeit fairly rough) estimate on the {global} gap of {$\HP$ and thus also $H$}.
\hfill $\Box$

\vspace{.2cm}

Let us conclude this section by briefly comparing the notion of a $\rho$-local gap of $H$ as defined here to another possibility of defining local spectral gaps, namely by using the LDOS (local density of states). It is defined by
$$
\mbox{\rm LDOS}(x,E)
\;=\;
\sum_n\delta(E_n-E)\,|\phi_n(x)|^2
\;\approx\;
\sum_n \frac{1}{\sigma \sqrt{2\pi}}\exp\big(-\tfrac{(E_n-E)^2}{2\sigma}\big)\,|\phi_n(x)|^2
\;,
$$
where the last expression is used as numerical approximation for some sufficiently small energy width $\sigma$. Here the sum runs over all eigenstates of the Hamiltonian. While standard in the physics literature for a long time, the formula on the r.h.s. has recently also been studied from a mathematical perspective in \cite{LLW} where it is called the window LDOS. One then sees a dip in $\mbox{\rm LDOS}(x,E)$ in a region of expected open local gap, but generally not a strict $0$, in particular, for systems exhibiting boundary or defect states. Instead, the advantage of the $\rho$-local gap is to provide a clean mathematical condition, which is also easily attainable by numerics, and, as will be seen below, particularly natural for the study of topological models.

%%%%%%%%%%%%%%%%%%%%%%%%%%%%%%%%%%%%%%%%%%%%%
\section{The spectral localizer and the local topological index}
\label{sec-SpecLocGap}

Let us focus on the case of even dimension $d$. {It is then well-known that the irreducible Clifford representation $\gamma_1,\ldots,\gamma_d$ admits another Clifford generator $\gamma_{d+1}$. In the following, we will choose a representation in which $\gamma_{d+1}$ is diagonal and given by the third Pauli matrix ({\it e.g.} Section 2.2.3 in \cite{PS}). Then $\Gamma=\one\otimes\gamma_{d+1}$ implements the chiral symmetry of the Dirac operator, namely $\Gamma D(x)\Gamma=-D(x)$. This implies
$$
D(x)
\;=\;
\begin{pmatrix}
0 & D_0(x)^* \\ D_0(x) & 0
\end{pmatrix}
\;,
\qquad
\Gamma
\;=\;
\begin{pmatrix}
\one & 0 \\ 0 & -\one
\end{pmatrix}
\;,
$$
for some $D_0(x)$ which is not selfadjoint. } Now the (even) spectral localizer of $H$ (in which a possible energy shift is absorbed) for tuning parameter $\kappa>0$ and at $x\in\RM^d$ is defined by
\begin{equation}
\label{eq-SpecLocDef}
L_{\kappa}(H,x)
\;=\;
\begin{pmatrix}
-\,H{\otimes\one_{\frac{d'}{2}}} & \kappa \,D_0(x)^*\\  \kappa\, D_0(x) & \,H{\otimes\one_{\frac{d'}{2}}}
\end{pmatrix}
\;=\;
-\,\HP\,\Gamma\,+\,\kappa\,D(x)
\;.
\end{equation}
Then $x$ is called the locality center of the localizer (or, in more physical terms, the probe site).  
{In order to define the finite volume restrictions, let us naturally extend the definitions of the partial isometries $\pi_\rho(x)$ and consequently $\one_\rho(x)=\pi_\rho(x)^*\pi_\rho(x)$ to the Clifford degrees of freedom by tensoring with the identity, and this without further notational distinction.} The finite volume spectral localizer is then denoted by 
$$
L_{\kappa,\rho}(H,x)
\;=\;
\pi_\rho(x)L_{\kappa}(H,x)\pi_\rho(x)^*
\;.
$$
Two quantities associated to the spectral localizer are of particular interest.

%%%%%%%%%%%%%%%%%%%%%%%%%
\begin{definition}
Let $H$ be weakly local. Then the localizer gap {\rm (}also referred to as the local topological protection gap{\rm )} is defined by
\begin{equation}
\label{eq-LocalTopGap}
\mu_{\kappa,\rho}(H,x)
\;=\;
\inf\,\spec\big(|L_ {\kappa,\rho}(H,x)|\big)\;,
\end{equation}
and, provided  that $\mu_{\kappa,\rho}(H,x)>0$,  the local topological index {\rm (}or local Chern marker{\rm )} by
\begin{equation}
\label{eq-LocalIndDef}
\LocInd_{\kappa,\rho}(H,x)\;=\;\frac{1}{2}\,\Sig(L_ {\kappa,\rho}(H,x))\;.
\end{equation}
\end{definition}
%%%%%%%%%%%%%%%%%%%%%%%%%

The main result of this work will provide sufficient conditions for a quantitative lower bound on the localizer gap and a stability statement on the local topological index. The proof will use a quantitative tapering estimate as the essential ingredient. A tapering function $F:\RM\to [0,1]$ is by definition continuously differentiable, even and satisfies $F(y)=0$ for $|y|\geq 1$ and $F(y)=1$ for $|y|\leq \frac{1}{2}$. Given $\rho>0$, its scaled version is then obtained by setting $F_\rho(y)=F(\frac{y}{\rho})$. The tapering estimate is then of the form
\begin{equation}
\label{eq-TaperingEst}
\|[F_\rho(D(x)),\HP](\imath\,\one +\tfrac{1}{\delta} D(x))^{-1}\|
\;\leq\;
\frac{C_F}{\rho}\,
\|[D(x),\HP](\imath\,\one +\tfrac{1}{\delta} D(x))^{-1}\|
\;,
\end{equation}
where $\delta>0$ is some parameter and $C_F$ is a bounding coefficient stemming from the behavior of the tapering function. Note that for $\delta=\infty$, the bound reduces to \eqref{eq-TaperingEstOld} with $C_F=8$. This case has been dealt with in earlier works \cite{LS2,DSW} which showed that $F$ can be chosen such that $C_F=8$, namely \eqref{eq-TaperingEstOld}. Appendix~\ref{app-tapering} provides a thorough discussion of the tapering estimate \eqref{eq-TaperingEst}. In particular, it proves that \eqref{eq-TaperingEst} also holds with $C_F=4.56 $ by choosing a different function $F$ and improving the estimates. Furthermore, it is also heuristically argued and numerically verified that for nearest neighbor hopping the constant in  \eqref{eq-TaperingEst} can be chosen to be $C_F\approx 2$. Now let us come to the main theorem. Admittedly, the statement is relatively technical on first sight, but below follow a series of remarks showing how it enfolds several interesting limit cases.

%%%%%%%%%%%%%%%%%%%%%%%%
\begin{theorem}
\label{theo-localGap}
Suppose that a weakly local Hamiltonian $H$ has a $\rho$-local gap $g_\rho=g_\rho(H,x)>0$ at $x$. Suppose that one can choose $\kappa$ such that both of the following inequalities hold:
\begin{equation}
\label{eq-kappaCond}
\frac{2\,g_\rho}{\rho}\;<\;\kappa\;\leq\; 
\frac{g_\rho^3}{\frac{1}{1-a-b^2}\big(
C_F\,
\|\HP R_{\kappa}\|\,
\,+\,
g_\rho\big)
\|[D(x),\HP]R_{\kappa}\|}
\end{equation}
where $a,b,c\geq 0$ with $1-a-b^2>0$ and
$$
R_{\kappa}
\;=\;
\big(\imath\,\one \,+\,\tfrac{c\,\kappa}{g_\rho} \, D(x)\big)^{-1}
\;,
\qquad
c^2\,=\,\tfrac{a}{1-a-b^2}
\;.
$$
Then for all  $\rho'\geq \rho$  one has 
$$
L_ {\kappa,\rho'}(H,x)^2
\;\geq\;
b^2\,g_\rho^2\,\one_{\rho'}(x)
\;,
$$  
which is an operator inequality on the finite dimensional Hilbert space $\Ran(\pi_{\rho'}(x))=\Ran(\one_{\rho'}(x))$. Otherwise stated, the localizer gap is then bounded below by the $\rho$-local gap of $H$ at $x$ as
\begin{equation}
\label{eq-TwoGaps}
\mu_{\kappa,\rho'}(H,x)
\;\geq\;
b\,g_\rho(H,x)
\;,
\qquad
\rho'\,\geq\, \rho
\;.
\end{equation}
Moreover, let $J\subset \RM^2$ be a simply connected region such that
\begin{align}
J
\;\subset\;
\big\{ (\kappa,\rho)\in\RM_>^2\,:\, 
(\kappa,\rho) \;\mbox{\rm satisfies }\eqref{eq-kappaCond}\big\}
\;.
\label{eq-Jdef}
\end{align}
Then \eqref{eq-TwoGaps} holds for $(\kappa,\rho)\in J$ and $(\kappa,\rho)\in J\mapsto \LocInd_{\kappa,\rho}(H,x)=\LocInd_{\kappa,\rho'}(H,x)$ is constant for $\rho'\geq\rho$.
\end{theorem}
%%%%%%%%%%%%%%%%%%%%%%%%

%%%%%%%%%%%%%%%
\begin{remark}
\label{rem-LocalityCond}
{\rm
The constants $a,b>0$ introduce some flexibility in the hypothesis \eqref{eq-kappaCond} as well as the implication \eqref{eq-TwoGaps}. For $a=0$, one has $R_\kappa=-\imath\,\one$ so that $\|\HP R_\kappa \|=\|H\|$ and $\|[D(x),\HP]R_\kappa\| =\|[D(x),\HP]\|$. Then \eqref{eq-kappaCond} reduces to
\begin{equation}
\label{eq-kappaCond3}
\frac{2\,g_\rho}{\rho}\;<\;\kappa\;\leq\; 
\frac{g_\rho^3}{\frac{1}{1-b^2}(C_F\,\|H\|+g_\rho)\,\|[D(x),\HP]\|}
\;,
\end{equation}
which is the same type of bound as in \cite{LS2,DSW}, except that only the local gap $g_\rho$ appears (which is always larger than or equal to the global gap, see Proposition~\ref{prop-LocalGap}). Let us note that $g_\rho$ is typically much smaller than $C_F\,\|H\|$ so that one can neglect $g_\rho$ in the denominator in \eqref{eq-kappaCond3}, as well as in \eqref{eq-kappaCond} in fact, and this will be done in the following discussion. In order to compare the constants in \eqref{eq-kappaCond3} with those of the prior works, let us choose $b=\frac{1}{2}$ so that \eqref{eq-TwoGaps} is the same outcome as in \cite{LS2,DSW}. In the latter references, the constant $\frac{1}{1-b^2}C_F=\frac{4}{3}C_F$ was $12$, which is reproduced by $C_F=9$, but $C_F=2$ implies that $\frac{4}{3}C_F=\frac{8}{3}$, notably an improvement by a factor $4.5$. If one is even satisfied with a gap of size $0.1\,g_\rho$, then the constant is merely $\frac{1}{1-b^2}C_F\approx 2$. This is fairly close to numerical observations, as it predicts that $\sim 20$ unit cells are needed in each lattice direction to guarantee a $\kappa$ that satisfies Eq.~\eqref{eq-kappaCond}, and these lattice sizes are observed to have stable localizer gaps that agree with the system's bulk spectral gap, resulting in stable local topological markers \cite{CL24}. On the other hand, for $a>0$ the relative operator norms (w.r.t.\ $D(x)$) appear in \eqref{eq-kappaCond}. In a homogeneous system, this does not lead to significantly smaller norms $\|\HP R_\kappa\|$ and $\|[D(x),\HP]R_\kappa\|$, but it does allow to show that modifications of the Hamiltonian far off the locality center $x$ do not modify the gap estimate considerably. Combined with the fact \eqref{eq-WeylType} that the gap $g_\rho=g_\rho(H,x)$ is also local and does even not depend on what $H$ is outside $B_\rho(x)$, this shows that the localizer gap and thus its half-signature only depend on local properties of the Hamiltonian near $x$, and this {\it locality property of the spectral localizer} comes in an explicit quantitative formulation. This is made more explicit in Section~\ref{sec-StabGap}.
}
\hfill $\diamond$
\end{remark}
%%%%%%%%%%%%%%%

%%%%%%%%%%%%%%%
\begin{remark}
\label{rem-LimitCond}
{\rm
The inequalities in \eqref{eq-kappaCond} can clearly be rewritten in the form
\begin{equation}
\label{eq-kappaCondAlt}
\frac{2}{\rho}\;<\;\frac{\kappa}{g_\rho}\;\leq\; 
\frac{g_\rho^2}{\frac{1}{1-a-b^2}\big(C_F\,\|\HP(\imath \,\one +c \frac{\kappa}{ g_\rho} D(x))^{-1}\|+g_\rho\big)\,\|[D(x),\HP](\imath \,\one +c\frac{\kappa}{ g_\rho} D(x))^{-1}\|}
\;.
\end{equation}
This stresses that $\kappa$ should naturally be related to the size of the local gap $g_\rho$.  In both \eqref{eq-kappaCond}  and \eqref{eq-kappaCondAlt}, the second inequality is non-linear in $\kappa$. Using the first inequality, it is, however, sufficient to impose the following stronger condition in which the r.h.s.\ is independent of $\kappa$
\begin{equation}
\label{eq-kappaCond2}
\frac{2\,g_\rho}{\rho}\;<\;\kappa\;\leq\; 
\frac{g_\rho^3}{\frac{1}{1-a-b^2}\big(C_F\,\|\HP(\imath \,\one +\frac{2c}{\rho} D(x))^{-1}\|+g_\rho\big)\|[D(x),\HP](\imath \,\one+\frac{2c}{\rho} D(x))^{-1}\|}
\;.
\end{equation}
Indeed, using spectral calculus and the first inequality in \eqref{eq-kappaCond}, one finds, setting $x=0$ for sake of simplicity,
\begin{align*}
\|\HP(\imath  \,\one+\tfrac{c\,\kappa}{g_\rho} D)^{-1}\|
&
\,\leq\,
\|\HP(\imath \,\one+\tfrac{2c}{\rho}\, D)^{-1}\|
\,
\|(\imath\,\one +\tfrac{2c}{\rho}\, D)(\imath  \,\one+\tfrac{c\,\kappa}{g_\rho}D)^{-1}\|
%\\
%&
\,\leq\,
\|\HP(\imath \,\one +\tfrac{2c}{\rho}\, D)^{-1}\|
\,,
\end{align*}
and similarly
$$
\|[D,\HP](\imath \,\one+\tfrac{c \kappa}{g_\rho} D)^{-1}\|
\;\leq\;
\|[D,\HP](\imath\,\one +\tfrac{2c}{\rho} D)^{-1}\|
\;,
$$
and these two bounds imply that the r.h.s.\ of \eqref{eq-kappaCond2} is smaller than or equal to the r.h.s.\ of \eqref{eq-kappaCond}, showing that \eqref{eq-kappaCond2} can be used as sufficient condition in Theorem~\ref{theo-localGap}. Fig.~\ref{fig:bounds} shows for a concrete model to what extend \eqref{eq-kappaCond} is stronger than \eqref{eq-kappaCond2}, and \eqref{eq-kappaCond2} is stronger than \eqref{eq-kappaCond3}, both  in a quantitative manner. 
}
\hfill $\diamond$
\end{remark}
%%%%%%%%%%%%%%%

%%%%%%%%%%%%
\begin{remark}
{\rm
In \eqref{eq-kappaCond} the gap $g_\rho(H,x)$ is local and does not depend on what $H$ is outside $B_\rho(x)$, however, the norms on the r.h.s.\ of \eqref{eq-kappaCond} are global. If one is merely interested to prove the lower bound  $\mu_{\kappa,\rho}(H,x)\geq b\,g_\rho(H,x)$ on the gap of $L_{\kappa,\rho}(H,x)$, one may replace the norms on the r.h.s.\ by $\|\HP_{\rho}\|$ and $\|[D_{\rho},\HP_{\rho}]\|$. However, an important aspect of Theorem~\ref{theo-localGap}  is that the gap remains open for all $\rho'\geq \rho$ and that the topological indices are the same at larger volumes (in fact, in Proposition~\ref{prop-IncreasingVolume} below it is stated that this larger volume need not be sets $B_{\rho'}(x)$, but merely needs to contain $B_\rho(x)$ and can be arbitrary otherwise). It is clear that making statements on $L_{\kappa,\rho'}(H,x)$ for $\rho'\geq \rho$ requires some information on the Hamiltonian $H$ in these larger volumes. According to Theorem~\ref{theo-localGap}, it is sufficient to have relative global operator norm estimates on $\HP$ and $[D,\HP]$ (relative w.r.t.\ $D$). The strength and importance of the new criterion \eqref{eq-kappaCond} compared to \eqref{eq-kappaCond3} from prior works is further discussed in Section~\ref{sec-StabGap}.
}
\hfill $\diamond$
\end{remark}
%%%%%%%%%%%%

%%%%%%%%%%%%%%%
\begin{remark}
{\rm
In conditions \eqref{eq-kappaCond}, the Hamiltonian $\HP$ and the commutator  $[D,\HP]$ are damped by the resolvent of $D$. In the realm of noncommutative geometry, {such} relative boundedness conditions  are natural and have been used \cite{KL} in order to assure the existence of the Kasparov product of unbounded operators. 
}
\hfill $\diamond$
\end{remark}
%%%%%%%%%%%%%%%

%%%%%%%%%%%%
\begin{remark}
{\rm
The bound \eqref{eq-Criterion2d} takes a slightly different form than \eqref{eq-kappaCond2}. In fact, in the proof of Theorem~\ref{theo-localGap} one also may replace $R_\kappa$ by $\hat{R}_\kappa=(\imath\,\one +\tfrac{c \kappa}{g_\rho} |D(x)|)^{-1}$. Then the criterion \eqref{eq-kappaCond} still holds with $R_\kappa$ replaced by $\hat{R}_\kappa$. Consequently in \eqref{eq-kappaCond2} one may also replace the two $D(x)$'s in the resolvents by $|D(x)|$. Furthermore, in dimension $d=2$ and for $x=0$, one has
$$
[D,\HP]
\;=\; 
\Big[
\begin{pmatrix}
0 & X_1-\imath X_2 \\ X_1+\imath X_2 & 0
\end{pmatrix}
,
\begin{pmatrix}
H & 0 \\ 0 & H
\end{pmatrix}
\Big]
\;=\;
\begin{pmatrix}
0 & [X_1-\imath X_2,H] \\ [H,X_1+\imath X_2] & 0
\end{pmatrix}
\;.
$$
Replacing this in \eqref{eq-kappaCond2} with $\hat{R}_\kappa$ and choosing $b=c=\frac{1}{2}$, namely $a=\frac{3}{20}$, one obtains \eqref{eq-Criterion2d}. 
}
\hfill $\diamond$
\end{remark}
%%%%%%%%%%%%%%%

\noindent {\bf Proof} of Theorem~\ref{theo-localGap}. For sake of notational simplicity, let us change coordinates such that $x=0$ and then denote $D=D(x)$, $L_{\kappa}=L_{\kappa}(H,x)$ and $L_{\kappa,\rho}=L_{\kappa,\rho}(H,x)$ as well as $g_\rho=g_\rho(H,x)$. Given the tapering functions $F$ and $F_\rho$ as above, one has $F_\rho(x)=0$ for $|x|\geq \rho$ and $F_\rho(x)=1$ for $|x|\leq \frac{\rho}{2}$. Finally set $F_\rho=F_\rho(D)$ which satisfies $F_\rho=F_\rho(|D|)$ and $[F_\rho,\Gamma]=0$. 

\vspace{.1cm}

First let us spell out the square of the localizer and then bound it from below by using the bound $\one_{\rho}\geq F_{\rho}^2$:
\begin{align*}
(L_{\kappa,\rho})^2
&
\;=\;
\pi_{\rho}\,
L_{\kappa}\,
\one_{\rho}\,
L_{\kappa}\,
\pi_{\rho}^*
\\
&
\;=\;
\pi_{\rho}\,
(\kappa D+\HP\Gamma)\,
\one_{\rho}\,
(\kappa D+\HP\Gamma)\,
\pi_{\rho}^*
\\
&
\;=\;
\kappa^2\, \pi_{\rho}D^2\pi_\rho^*
\,+\,
\pi_{\rho}\HP\one_\rho \HP\pi_\rho^*
\,+\,
\kappa\,\pi_{\rho}[D,\HP]\Gamma \pi_\rho^*
\\
&
\;\geq\;
\kappa^2\, \pi_{\rho}D^2\pi_\rho^*
\,+\,
\pi_{\rho}\HP F^2_\rho \HP\pi_\rho^*
\,+\,
\kappa\,\pi_{\rho}[D,\HP]\Gamma \pi_\rho^*
\\
&
\;=\;
\kappa^2\, \pi_{\rho}D^2\pi_\rho^*
\,+\,
\pi_\rho F_{\rho} (\HP)^2 F_{\rho}\pi_\rho^*\,+\,
\pi_\rho\big(
[\HP,F_{\rho}] F_{\rho} \HP \,+\,F_{\rho}\HP[F_{\rho},\HP]
\,+\,
\kappa\,[D,\HP]\Gamma\big) \pi_\rho^*
\;.
\end{align*}
The first summand is bounded below as follows, for a constant $a\in[0,1)$,
\begin{align}
\kappa^2\, \pi_{\rho}D^2\pi_\rho^*
&
\;=\;
a\,\kappa^2\, \pi_{\rho}D^2\pi_\rho^*
\,+\,(1-a)\,\kappa^2\, \pi_{\rho}D^2\pi_\rho^*
\nonumber
\\
&
\;\geq\;
a\,\kappa^2\, \pi_{\rho}D^2\pi_\rho^*
\,+\,(1-a)\,\kappa^2\, \pi_{\rho}D(\one-F_\rho^2)D\pi_\rho^*
\label{eq-Dsplit}
\\
&
\;\geq\;
a\,\kappa^2\, \pi_{\rho}D^2\pi_\rho^*
\,+\,
(1-a)\,g_\rho^2\,{\pi_{\rho}(\one-F_\rho^2)\pi_{\rho}^*}
\;,
\nonumber
\end{align}
because the bound  holds for spectral parameters in {$[\frac{1}{2}\rho,\rho]$} due to the first bound in \eqref{eq-kappaCond} and since $\one-F_\rho^2\leq \one$, while it holds trivially on $[0,\frac{1}{2}\rho]$. For the second summand, the definition of the local gap gives
$$
F_{\rho} (\HP)^2 F_{\rho}
\;\geq\; 
g^2_\rho F_\rho^2
\;.
$$
Replacing these estimates in the above yields
\begin{align*}
(L_{\kappa,\rho})^2
&
\;\geq\;
a\,\kappa^2\, \pi_{\rho}D^2\pi_\rho^*
\,+\,
(1-a)\,g_\rho^2\,\pi_{\rho}(\one-F^2_\rho)\pi_\rho^*
\,+\,
g^2_\rho \pi_\rho F_\rho^2\pi_\rho^*
\,+\,
\pi_\rho B \pi_\rho^*
\;,
\end{align*}
where 
$$
B
\;=\;
[\HP,F_{\rho}] F_{\rho} \HP \,+\,F_{\rho}\HP[F_{\rho},\HP]
\,+\,
\kappa\,[D,\HP]\Gamma
\;.
$$
Now as $(1-a)(1-F^2)+F^2\geq (1-a)$ for $F\in[0,1]$, and, provided  $b^2<1-a$ and with $R=R_\kappa$ being the invertible operator introduced in the statement of the theorem,
\begin{align*}
(L_{\kappa,\rho})^2
&
\;\geq\;
a\,\kappa^2\, \pi_{\rho}D^2\pi_\rho^*
\,+\,
(1-a)\,g_\rho^2\,\one_{\rho}
\,+\,
\pi_\rho B \pi_\rho^*
\\
&
\;=\;
b^2\,g_\rho^2\,\one_{\rho}
\,+\,
\pi_\rho \big((1-a-b^2)\,g_\rho^2\,\one\,+\,a\,\kappa^2\,D^2  \,+\,B\big)\pi_\rho^*
\\
&
\;=\;
b^2\,g_\rho^2\,\one_{\rho}
\,+\,
(1-a-b^2)\,g_\rho^2\,\pi_\rho \big((R^{-1})^*R^{-1}  \,+\,\tfrac{1}{1-a-b^2}\,\tfrac{1}{g_\rho^2}\,B\big)\pi_\rho^*
\\
&
\;=\;
b^2\,g_\rho^2\,\one_{\rho}
\,+\,
(1-a-b^2)\,g_\rho^2\,\pi_\rho (R^{-1})^*\big(\one  \,+\,\tfrac{1}{1-a-b^2}\,\tfrac{1}{g_\rho^2}\,R^*BR\big)R^{-1}\pi_\rho^*
\;.
\end{align*}
From this one infers
$$
{\|R^*BR\|}\,\leq\,(1-a-b^2)\,g_\rho^2
\quad \Longrightarrow\quad
(L_{\kappa,\rho})^2
\,\geq\,b^2\,g_\rho^2\,\one_\rho
\;.
$$
Hence let us bound, using $[F_\rho,R]=0$, $\|F_\rho\|\leq 1$, $\|\Gamma\|=1$ and the bound \eqref{eq-TaperingEst}: 
\begin{align*}
\| R^*  BR\|
&
\;\leq\;
\|R^*[\HP,F_{\rho}] F_{\rho} \HP R\|
\,+\,
\|R^*F_{\rho}\HP[F_{\rho},\HP]R\|
\,+\,
\kappa\,\|R^*[D,\HP]\Gamma R\|
% \\ & \;\leq\; 2\, \|\HP R\|\,\|[F_{\rho},\HP]R\| \,+\, \kappa\,\|R^*[D,\HP]\Gamma R\|
\\
&
\;\leq\;
2\,
\|\HP R\|\,\|[F_{\rho},\HP]R\|
\,+\,
\kappa\,\|[D,\HP]R\|
\\
&
\;\leq\;
2\,C_F\,\tfrac{1}{\rho}\,
\|\HP R\|\,\|[D,\HP]R\|
\,+\,
\kappa\,\|[D,\HP]R\|
\\
&
{\;=\;}
\big(
2\,C_F\,\tfrac{1}{\rho}\,
\|\HP R\|\,
\,+\,
\kappa\big)
\|[D,\HP]R\|
\\
&
\;\leq\;
\kappa\,\tfrac{1}{g_\rho}\big(
C_F\,
\|\HP R\|\,
\,+\,
g_\rho\big)
\|[D,\HP]R\|
\;,
\end{align*}
where in the last step the first inequality of \eqref{eq-kappaCond} was used again. The latter quantity is smaller than or equal to $(1-a-b^2)g_\rho^2$ provided that the second inequality of \eqref{eq-kappaCond} of holds.

\vspace{.1cm}

Next, let us show that
$$
\Sig\left(L_{\kappa,\rho}\right)
\;=\;
\Sig\left(L_{\kappa',\rho'}\right)
\;,
$$
for $(\kappa,\rho)\in J$ and $(\kappa',\rho')\in J$ with $\rho'\geq\rho$. As $L_{\kappa,\rho}$ is continuous in $\kappa$ and the gap remains open by the above argument,  it is sufficient to consider the case $\kappa=\kappa'$.  For this purpose, an adaption of the  homotopy argument given in \cite{LS2,DSW} will be used. For $\lambda\in[0,1]$, set
$$
L_{\kappa,\rho,\rho'}(\lambda)
\;=\;
\pi_{\rho'}\,
(\kappa D+F_{\lambda,\rho,\rho'} \HP F_{\lambda,\rho,\rho'}\Gamma)\,
\pi_{\rho'}
\;,
\qquad
F_{\lambda,\rho,\rho'}
\;=\;
(1-\lambda)\,\one_{\rho'}\,+\,\lambda\,F_\rho
\;.
$$
Note that then $L_{\kappa,\rho,\rho'}(0)=L_{\kappa,\rho'}$ so that the aim is to show
$$
\Sig\left(L_{\kappa,\rho,\rho}(0)\right)
\;=\;
\Sig\left(L_{\kappa,\rho,\rho'}(0)\right)
\;,
$$
For this purpose, let us first prove that the gap of $\lambda\in[0,1]\mapsto L_{\kappa,\rho,\rho'}(\lambda)$ is open as long as \eqref{eq-kappaCond} holds. This can be done by a minor modification of the above argument. Let us start out as following, using the bound $\one_{\rho}\geq F_{\rho}^2$ such that
\begin{align*}
L_{\kappa,\rho,\rho'}(\lambda)^2
%&
%\;=\;
%\pi_{\rho'}\,
%(\kappa D+F_{\lambda,\rho,\rho'} \HP\Gamma F_{\lambda,\rho,\rho'})\,
%(\kappa D+F_{\lambda,\rho,\rho'} \HP\Gamma F_{\lambda,\rho,\rho'})\,
%\pi_{\rho'}^*
%\\
&
\;=\;
\kappa^2\, \pi_{\rho'}D^2\pi_{\rho'}^*
\,+\,
\pi_{\rho'}F_{\lambda,\rho,\rho'} \HP F_{\lambda,\rho,\rho'}^2 \HP F_{\lambda,\rho,\rho'}\pi_{\rho'}^*
\,+\,
\kappa\,\pi_{\rho'}F_{\lambda,\rho,\rho'}[D,\HP]\Gamma F_{\lambda,\rho,\rho'} \pi_{\rho'}^*
\\
&
\;\geq\;
\kappa^2\, \pi_{\rho'}D^2\pi_{\rho'}^*
\,+\,
\pi_{\rho'}F_{\lambda,\rho,\rho'} \HP F_{\rho}^2 \HP F_{\lambda,\rho,\rho'}\pi_{\rho'}^*
\,+\,
\kappa\,\pi_{\rho'}F_{\lambda,\rho,\rho'}[D,\HP]\Gamma F_{\lambda,\rho,\rho'} \pi_{\rho'}^*
\\
&
\;=\;
\kappa^2\, \pi_{\rho'}D^2\pi_{\rho'}^*
\,+\,
\pi_{\rho'}  F_{\lambda,\rho,\rho'}F_\rho (\HP)^2 F_\rho F_{\lambda,\rho,\rho'} \pi_{\rho'}^*
\\
&
\;\;\;\;\;\;
\,+\,
\pi_{\rho'}
F_{\lambda,\rho,\rho'} \big(
[\HP,F_{\rho}]F_{\rho} \HP \,+\,F_{\rho}\HP[F_{\rho},\HP]
\,+\,
\kappa\,[D,\HP]\Gamma\big)F_{\lambda,\rho,\rho'} \pi_{\rho'}^*
\;.
\end{align*}
Instead of \eqref{eq-Dsplit}, the first summand will be estimated by
\begin{align*}
\kappa^2\, \pi_{\rho}D^2\pi_\rho^*
%&
%\;=\;
%\tfrac{1}{4}\,\kappa^2\, \pi_{\rho}D^2\pi_\rho^*
%\,+\,\tfrac{3}{4}\,\kappa^2\, \pi_{\rho}D^2\pi_\rho^*
%\nonumber
%\\
&
\;\geq\;
a\,\kappa^2\, \pi_{\rho}D^2\pi_\rho^*
\,+\,(1-a)\,\kappa^2\, \pi_{\rho}D(\one-F_\rho^4)D\pi_\rho^*
%\label{eq-Dsplit}
\\
&
\;\geq\;
a\,\kappa^2\, \pi_{\rho}D^2\pi_\rho^*
\,+\,
(1-a)\,g_\rho^2\,(\one-F_\rho^4)
\;.
\nonumber
\end{align*}
From this point on, the same argument as above shows that $L_{\kappa,\rho,\rho'}(\lambda)^2\geq b\,g_\rho^2\,\one_{\rho'}$. As $\lambda\in[0,1]\mapsto L_{\kappa,\rho,\rho'}(\lambda)$ is a continuous path in the invertibles, it suffices to prove 
\begin{equation}
\label{eq-SigEqual}
\Sig\left(L_{\kappa,\rho,\rho}(1)\right)
\;=\;
\Sig\left(L_{\kappa,\rho,\rho'}(1)\right)
\;.
\end{equation}
Consider 
$$
L_{\kappa,\rho,\rho'}(1)
\;=\;
\kappa\pi_{\rho'}D\pi_{\rho'}^{*}
+\pi_{\rho'}F_{\rho}(D)\,(-\HP\,\Gamma)\, F_{\rho}(D)\pi_{\rho'}^{*}
\;.
$$
{Now let $\pi_{\rho,\rho'}^*$ be the surjective partial isometry on the states supported by the ring $B_{\rho'}\setminus B_\rho$, namely $\pi_{\rho'}=\pi_{\rho,\rho'}\oplus \pi_{\rho}$ and the range of $\pi_{\rho,\rho'}^*$ is $(\Hh\oplus\Hh)_{\rho'}\ominus (\Hh\oplus\Hh)_{\rho}$.  Furthermore, $D$ leaves all the above subspaces invariant so that $D\pi_{\rho'}^{*}=D\pi_{\rho,\rho'}^{*}\oplus D\pi_{\rho}^{*}$ and also $F_{\rho}(D)\pi_{\rho'}^{*}=(F_{\rho}(D)\pi_{\rho}^{*})\oplus(0\cdot\pi_{\rho,\rho'}^*)$. Together one concludes
}
$$
L_{\kappa,\rho,\rho'}(1)
\;=\;
L_{\kappa,\rho,\rho}(1)\oplus\,\pi_{\rho,\rho'}\,\kappa\,D\,\pi_{\rho,\rho'}^{*}
\;.
$$
{The signature of $\pi_{\rho,\rho'}\,D\,\pi_{\rho,\rho'}^{*}$ vanishes by Sylvester's theorem because $\Gamma$ commutes with $\pi_{\rho,\rho'}^{*}$ and therefore the chirality of $D$ shows $\Gamma \pi_{\rho,\rho'}\,D\,\pi_{\rho,\rho'}^{*}\Gamma=-\pi_{\rho,\rho'}\,D\,\pi_{\rho,\rho'}^{*}$. Hence} \eqref{eq-SigEqual} follows.
\hfill $\Box$

\vspace{.2cm}

The final statement of Theorem~\ref{theo-localGap}  is that, if $(\kappa,\rho)\in J$, the local index $\LocInd_{\kappa,\rho}(H,x)$ is equal to $\LocInd_{\kappa,\rho'}(H,x)$ for arbitrary large $\rho'\geq\rho$. Actually it is possible to replace the set $B_{\rho'}(x)$ by any arbitrary set containing  $B_{\rho}(x)\cap\ZM^d$, for which the initial estimate is supposed to hold. More concretely, consider a subset $\Lambda\subset\ZM^d$ and let $\pi_\Lambda:\ell^2(\ZM^d,\CM^L)\to\ell^2(\Lambda,\CM^L)$ denote the associated surjective partial isometry. Then for an operator $A$ on $\ell^2(\ZM^d,\CM^L)$ its restriction to $\Lambda$ is $A^\Lambda=\pi_\Lambda A\pi_\Lambda^*$. The spectral localizer on $\Lambda$ is then denoted by $L^\Lambda_{\kappa}(H,x)=L_{\kappa}(H,x)^\Lambda$, and its gap and half-signature by $\mu_{\kappa}^\Lambda(H,x)$ and $\LocInd_{\kappa}^\Lambda(H,x)=\frac{1}{2}\,\Sig(L^\Lambda_ {\kappa}(H,x))$. If in the proof of Theorem~\ref{theo-localGap} one replaces $B_{\rho'}(x)$ by $\Lambda$, then one obtains the following statement:

%%%%%%%%%%%%%%%%%%%%%%%%
\begin{proposition}
\label{prop-IncreasingVolume}
Suppose that a weakly local Hamiltonian $H$ has a $\rho$-local gap $g_\rho(H,x)$ at $x$ and let $\kappa>0$ be such that the inequalities \eqref{eq-kappaCond} hold. If $\Lambda\subset\ZM^d$ is a finite subset satisfying $B_\rho(x)\cap\ZM^d\subset\Lambda$, then $L^\Lambda_ {\kappa}(H,x)^2\geq b^2g_\rho(H,x)^2\,\one_\Lambda$ and the topological index in $\Lambda$ is equal to that in $B_\rho(x)$, namely 
\begin{equation}
\label{eq-IncreaseVolume}
\mu_{\kappa}^\Lambda(H,x)
\;\geq\; 
b\,g_{\rho}(H,x)
\;,
\qquad
\LocInd_{\kappa}^\Lambda(H,x)
\;=\;
\LocInd_{\kappa,\rho}(H,x)\;.
\end{equation}
\end{proposition}
%%%%%%%%%%%%%%%%%%%%%%%%

%%%%%%%%%%%%%%%%%%%%%%%%%%%%%%%%%%%%%%%%%%%%%
\section{Stability of the localizer gap}
\label{sec-StabGap}

This section is about proving stability statements on the localizer gap $\mu_{\kappa}^\Lambda(H,x)$ under a weakly local perturbation $W=W^*$ of the Hamiltonian which is supported near a point $y$ that is distant from $x$. This discussion will highlight the strength of the new criterion \eqref{eq-kappaCond} when compared with the hypothesis of the prior works \cite{LS2,DSW}. Indeed, a naive approach is to simply view $W$ as a perturbation in $L_{\kappa,\rho}(H+W,x)$ and conclude by a Weyl-type estimate that the localizer gap remains open if $\|W\|<\mu_{\kappa,\rho}(H,x)$. However, if $\supp(W)\cap B_\rho(x)=\emptyset$, namely the support of $W$ is outside of $B_\rho(x)$, and the Hamiltonian is {finite range}, one clearly has $L_{\kappa,\rho}(H+W,x)=L_{\kappa,\rho}(H,x)$ so that $\mu_{\kappa,\rho}(H+W,x)=\mu_{\kappa,\rho}(H,x)$ and $\LocInd_{\kappa,\rho}(H+W,x)=\LocInd_{\kappa,\rho}(H,x)$. Nevertheless, the criterion \eqref{eq-kappaCond3} from earlier works \cite{LS2,DSW} will not guarantee this stability for large $W$ even if the local gap stability \eqref{eq-WeylType}  is used, simply because there are operators norms $\|H+W\|$ and  $\|[{H'+W'},D]\|$ on the r.h.s. of \eqref{eq-kappaCond3} which grow with the size of $W$ {(here again $W'=W\otimes\one_{d'}$ as for $H'$)}. Let us now show the improvement based on the criterion \eqref{eq-kappaCond2} with $b=c=\frac{1}{2}$, namely $a=\frac{3}{20}$, for a perturbation $W$ that is small near $x$. More precisely, as an input, it will be supposed that $\|W(\one+|X-y|)\|<\infty$ for some $y\in\RM^d$ which can be thought of as the localization center of $W$, namely $\|W(\one+|X-y|)\|\approx \|W\|$.  By \eqref{eq-WeylType}, one then knows that the $\rho$-local gap $g_{\rho}(H+W,x)$ is roughly the same as $g_\rho(H,x)$, provided $x-y$ is sufficiently large. {For sake of notational simplicity, we will simply denote $g_\rho=g_{\rho}(H+W,x)$.} Furthermore, the relative operator norms can be bounded as follows:
\begin{align*}
\|  (\HP+{W'}) & (\imath\,\one+\tfrac{1}{\rho}\,D(x))^{-1}\|
\;\leq\;
\|\HP(\imath\,\one+\tfrac{1}{\rho}\,D(x))^{-1}\|
\,+\,
\|{W'}(\imath\,\one+\tfrac{1}{\rho}\,D(x))^{-1}\|
\\
&
\;\leq\;
{\|H\|}
\,+\,
\|W(\one+|X-y|)\|\,\|(\one+|X-y|{\otimes\one_{d'}})^{-1}(\imath\,\one+\tfrac{1}{\rho}\,D(x))^{-1}\|
\;,
\end{align*}
{where in the first summand the resolvent was simply discarded because it is irrelevant for the main point of this section.  Let us compute the square of the last factor with the $C^*$-equation $\|A\|^2=\|A^*\|^2=\|AA^*\|$ and the identity $D(x)^*D(x)=|X-x|^2\otimes\one_{d'}$, and then estimate it using spectral calculus of $X$:
\begin{align*}
\|(\one & +|X-y|\otimes\one_{d'})^{-1}(\imath\,\one+\tfrac{1}{\rho}\,D(x))^{-1}\|^2
\\
&
\;=\;
\|(\one+|X-y|)^{-1}(\one+\tfrac{1}{\rho^2}|X-x|^2)^{-1}(\one+|X-y|)^{-1}\|
\\
&
\;\leq\;
\rho^2\,
\|(\one+|X-y|^2)^{-1}(\one+|X-x|^2)^{-1}\|
\\
&
\;=\;
\rho^2\,
\sup_{n\in\ZM^d}\,
\frac{1}{1+|n-y|^2}\,\frac{1}{1+|n-x|^2}
\\
&
\;\leq\;
\rho^2\,\frac{2}{1+|x-y|^2}
\;.
\end{align*}
Replacing the square root of this expression in the above estimate, one concludes that
\begin{align*}
\|  (\HP+W') (\imath\,\one+\tfrac{1}{\rho}\,D(x))^{-1}\|
\;\leq\;
\|H\|
\,+\,
\|W(\one+|X-y|)\|\,\rho\,\frac{4}{1+|x-y|}
\;.
\end{align*}
}
In a similar manner, one can bound $\|[D,\HP+{W'}](\imath\,\one+\tfrac{1}{\rho}\,D)^{-1}\|$. Setting {furthermore $W'(y)=W'(\one+|X-y|)$}, then \eqref{eq-kappaCond2} shows that the slightly stronger condition on $\kappa$ given by
$$
\frac{2\,{g_{\rho}} }{\rho}\,<\,\kappa\,<\, 
\frac{{(g_{\rho})^3}}{\frac{5}{3}\,\big(C_F\|H\|+\tfrac{{2}C_F\rho}{1+{|x-y|}}\,\|{W'}(y)\|+g_\rho
\big)\big(\|[D,\HP]\|+\tfrac{4\rho}{1+{|x-y|}}\,\|[{W'}(y),D]\|
\big)}
$$
is sufficient to conclude by Theorem~\ref{theo-localGap} and Proposition~\ref{prop-IncreasingVolume} that $\mu^\Lambda_\kappa(H+sW,x)\geq \frac{1}{2}\,g_\rho(H+W,x)$ for all $s\in[0,1]$. Hence by a homotopy in $s$  one has $\LocInd^\Lambda_\kappa(H+W,x)=\LocInd^\Lambda_\kappa(H,x)$ so that the local index does not change. Clearly the above condition is much weaker than prior criteria because the norms of $W$ and $[D,W]$ are divided by {$|x-y|$}. In other words, distant perturbations do not alter the local topological protection gap and the local topological index.

%%%%%%%%%%%%%%%%%%%%%%%%%%%%%%%%%%%%%%%%%%%%%
\section{Numerical illustration of local gaps and indices}
\label{sec-Numerics}

To illustrate the local behavior of the $\rho$-local gap $g_\rho$ and the improvements on finding a suitable parameter $\kappa$ for use in the spectral localizer framework, let us consider lattices and heterostructure containing the Haldane model \cite{Hal}:
\begin{align}
H_{\textrm{Hal}}  
\;=\; 
&
\sum_{n_A,n_B} 
 \big(M\,|n_A\rangle\langle n_A|\,-\,M\,|n_B\rangle\langle n_B|\big) 
 \,-\, 
 t \sum_{\langle n_A,m_B\rangle} 
 \big(|n_A\rangle\langle m_B|\,+\,|m_B\rangle\langle n_A|\big) 
\nonumber
\\
& \,-\,
 t_c\sum_{\alpha=A,B} \sum_{\langle\!\langle n_\alpha,m_\alpha\rangle\!\rangle}  
 \big(e^{\imath\phi(n_\alpha,m_\alpha)}\, |n_\alpha\rangle\langle m_\alpha|\,+\,e^{-\imath\phi(n_\alpha,m_\alpha)}\,|m_\alpha\rangle\langle n_\alpha|\big) 
 \,,
 \label{eq-haldaneH}
\end{align}
{where
$$
\phi(n_\alpha,m_\alpha)
\;=\;
\left\{
\begin{array}{cc}
\phi\;,
&
\mbox{\rm if }\alpha=A\;,
\\
-\phi\;,
&
\mbox{\rm if }\alpha=B\;.
\end{array}
\right.
$$
}
The Haldane lattice is a honeycomb lattice with nearest-neighbor couplings $t$, on-site energies $\pm M$ on the two sublattices of the honeycomb lattice, and next-nearest neighbor couplings with amplitude $t_c$ and phase $\pm \phi$. For a broad range of choices $M$, $t_c$, and $\phi$ such that $|M| < 3\sqrt{3} t_c \sin(\phi)$, this system is known to be in a non-trivial phase with Chern number $|\textrm{Ch}| = 1$.

\vspace{.2cm}

\begin{figure*}[t]
    \centering
    \includegraphics[width=\columnwidth]{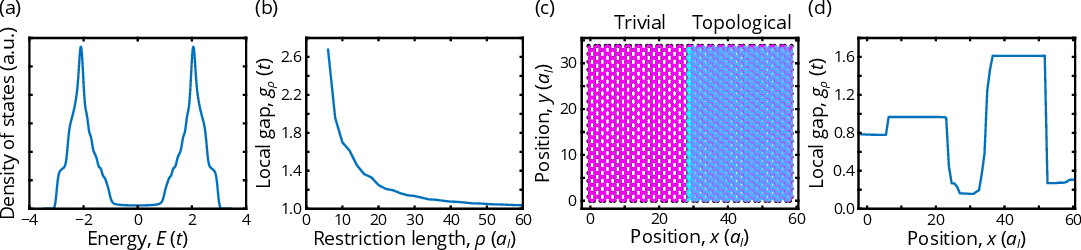}
    \caption{Demonstration that the local gap $g_\rho$ behaves as expected. (a) Density of states for a Haldane lattice with $t_c = t/2$, $\phi = \pm \pi/2$, and $M = 0$. (b) Local gap $g_\rho$ as a function of the restriction length $\rho$ for the same Haldane lattice of size $80 \times 80$ with the restriction region's center coinciding with the lattice's. The site-to-site spacing is $a_l$. (c) Schematic of a heterostructure formed by massive graphene with $t_c = 0$, $\phi = 0$, and $M = (\sqrt{3}/2) t$ (left) and the same topological Haldane lattice (right). (d) Local gap for $\rho = 12a_l$ as the restriction region's center $x$ is varied across the lattice.}
    \label{fig:gapDemo}
\end{figure*}

Figure~\ref{fig:gapDemo} demonstrates that the behavior of the local gap from Definition~\ref{def-LocalGapDirichlet} has the desired intuitive properties. In particular, Figs.~\ref{fig:gapDemo}(a) and (b) show that as $\rho$ increases without touching a boundary or interface, the local gap converges to the bulk spectral gap. Similarly, Figs.~\ref{fig:gapDemo}(c) and (d) show that as the restriction region's center is varried across a lattice's boundaries and interfaces where there are localized states that exist within the bulk spectral gaps of each of the heterostructure's constituent materials, the local gap closes.

\vspace{.2cm}

\begin{figure*}[t]
    \centering
    \includegraphics{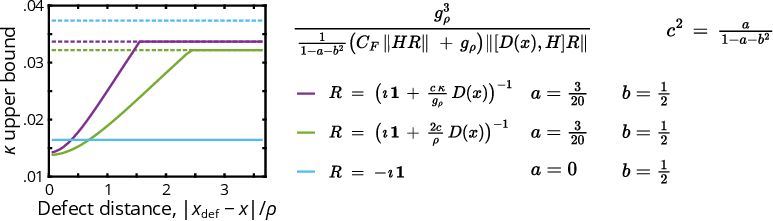}
    \caption{Behavior of the upper bound on $\kappa$ from Theorem~\ref{theo-localGap} for a large Haldane lattice with $100 \times 100$ sites as a single (large) defect is moved from the system's center to its boundary. The total system Hamiltonian $H = H_{\textrm{Hal}} + W(x_{\textrm{def}})$, where $W(x_{\textrm{def}})$ has only a single non-zero entry along its diagonal corresponding to the site at location $x_{\textrm{def}}$ with value $7t$. Solid lines show the bounds behavior in the presence of the defect, while dashed lines consider only the clean system. Here, $t_c = t/2$, $\phi = \pm \pi/2$, $M = 0$, and we use $\rho = 20a_l$ where $a_l$ is the site-to-site spacing. The local gap is calculated at the lattice's center $x$. Green lines show the r.h.s. of Eq.~\eqref{eq-kappaCond} with $a = 3/20$ and $b = 1/2$. Purple lines show the r.h.s. of Eq.~\eqref{eq-kappaCond3} with $a = 3/20$ and $b = 1/2$. Blue lines show the r.h.s. of Eq.~\eqref{eq-kappaCond} with $a = 0$ and $b = 1/2$. $\kappa = 0.2t/a_l$ is used for the relevant bound (green).}
    \label{fig:bounds}
\end{figure*}

Figure~\ref{fig:bounds} shows the advantage of the bounds on $\kappa$ in Theorem~\ref{theo-localGap} when adding defects. In particular, a strong defect in a distant portion of a system will strongly influence $\Vert H \Vert$, whereas the new bounds decrease the influence of any such defect when it is far away from the location where the system's local properties are being considered. Indeed, for a sufficiently distant defect, the bound in Theorem~\ref{theo-localGap} returns to its value for the corresponding unperturbed system, while noting the need for a non-zero $a$ in the r.h.s. of Eq.~\eqref{eq-kappaCond}. Systems that are globally ordered still obtain the best bounds on $\kappa$ because one can set $a=0$ in Eq.~\eqref{eq-kappaCond}.

\vspace{.2cm}

\begin{figure*}[t]
    \centering
    \includegraphics{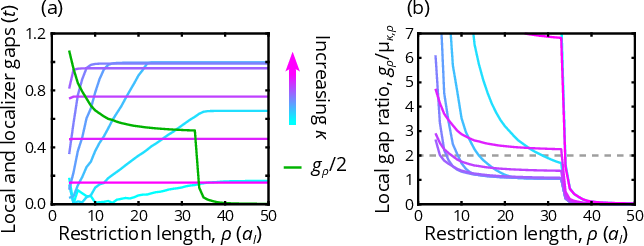}
    \caption{(a) Comparison of the local gap $g_\rho$ against the localizer gap $\mu_{\kappa,\rho}$ for different values of $\kappa$ (in units of $t/a_l$) as the restriction radius is increased, $\kappa = [0.005, 0.02, 0.05, 0.1, 0.2, 0.5, 1, 1.5, 2](t/a_l)$ (cyan to magenta in increasing order) and $g_\rho/2$ (green). (b) Ratio $g_\rho/\mu_{\kappa,\rho}$ for the same choices of $\kappa$. Dashed gray line corresponds to $g_\rho/\mu_{\kappa,\rho} = 2$, corresponding to $b = 1/2$. The system Hamiltonian is given by a Haldane lattice adjacent to a massive graphene lattice nearly identical to Fig.~\ref{fig:gapDemo}(c), except it is $90 \times 90$ sites in total size. The center of the restriction region is taken to be in the center of the topological domain, such that $\rho \sim 33a_l$ corresponds to the restriction region beginning to include both the lattice boundary and the interface between the two constituent materials.}
    \label{fig:muGap}
\end{figure*}

However, even the improved bounds in Theorem~\ref{theo-localGap} are not generally optimal, as shall be illustrated in the following. This implies that the local index $\LocInd_{\kappa,\rho}(H,x)$ may still be useful even for choices of $\rho$ and $\kappa$ that do not satisfy Eq.~\eqref{eq-kappaCond}. In particular, the region where the numerically computed quotient $\frac{g_\rho(H,x)}{\mu_{\kappa,\rho}(H,x)}$ is less than $\frac{1}{b}$ in Eq.~\eqref{eq-TwoGaps} for a given $\kappa$ is expected to yield accurate results for the calculation of any local index.

\vspace{.2cm}

Another important consequence of Eq.~\eqref{eq-TwoGaps} is that for larger volumes $\rho'\geq\rho$ the localizer gap satisfies $\mu_{\kappa,\rho'}(H,x)\geq b\,g_\rho(H,x)$. Hence, if $g_\rho(H,x)>0$, the function $\rho'\to \mu_{\kappa,\rho'}(H,x)$ remains positive and actually converges to some positive value as $\rho'\to\infty$. For all such $\rho'$ the local index can still {safely} be computed as the half-signature of $L_{\kappa,\rho'}(H,x)$. All this holds even if the boundary of $B_{\rho'}(x)$ reaches a topological interface. For such a $\rho'$ one then has $g_{\rho'}(H,x)\approx 0$ and the quotient $\frac{g_{\rho'}(H,x)}{\mu_{\kappa,\rho'}(H,x)}$ then converges to $0$. These features are illustrated in Fig.~\ref{fig:muGap} for different choices of $\kappa$. 

%%%%%%%%%%%%%%%%%%%%%%%%%%%%%%%%%%%%%%%%%%%%%
\section{Local gap distribution in Anderson localized phase}
\label{sec-AndLoc}

This section considers random topological Hamiltonians of the type
\begin{equation}
H_\omega(\lambda)\;=\;H\,+\,\lambda\,V_\omega \label{eq:Hdisorder}
\;, 
\end{equation}
where $H$ is a gapped possibly topological homogeneous Hamiltonian and $V_\omega=\sum_{n\in\ZM^d}v_n\,|n\rangle\langle n|$ is a random potential associated to a configuration $\omega=(v_n)_{n\in\ZM^d}$ given by independent real random variables $v_n$ drawn from the interval $[-\frac{1}{2},\frac{1}{2}]$ according to the uniform distribution. It is well-known that the spectrum of $H_\omega(\lambda)$ is almost surely the same, that it grows with $\lambda$, and that one expects most of it to be Anderson localized, namely the spectrum consists of a dense set of eigenvalues with exponentially localized eigenfunctions (this is proved only for the band edges and at high disorder, see \cite{AW} for a review). Moreover, as $\lambda$ increases, the gap of $H$ closes and  fills with such Anderson localized states. It is known \cite{BES,SSt} that the Chern numbers are well-defined and constant in an interval of dynamical Anderson localization. Another important piece of information that will be used here is that the eigenvalues and the localization centers of their eigenfunctions are, after appropriate rescaling, given by two independent Poisson processes with intensity specified by the density of states \cite{GK}. While this latter result does not strictly cover the case of a topological Hamiltonian and its square, we will freely use this information in the heuristic arguments of this section.

\vspace{.2cm}

\begin{figure*}[t]
    \centering
    \includegraphics[width=\columnwidth]{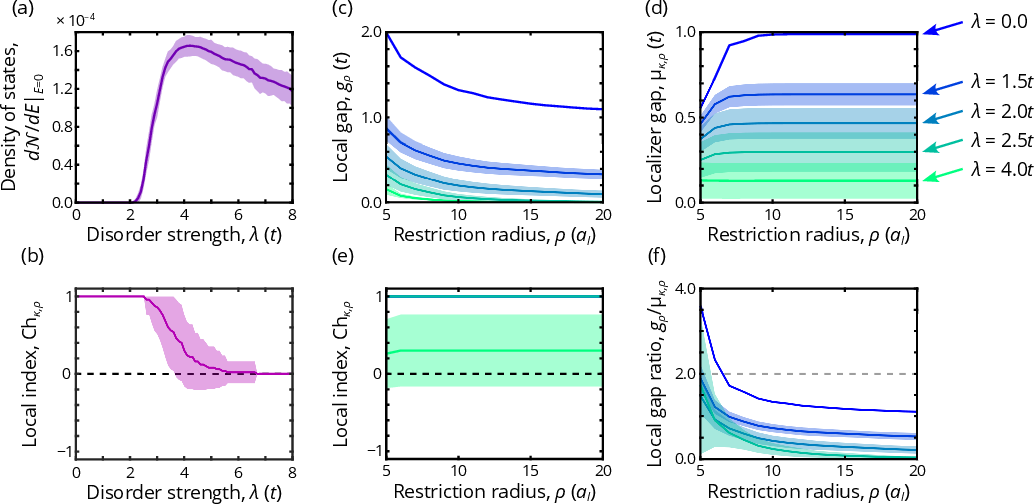}
    \caption{Analysis of the local gap behavior as disorder is added to the Haldane lattice with $t_c = t/2$, $\phi = \pm \pi/2$, and $M = 0$. (a) Density of states at $E = 0$ as the disorder strength is increased for 50 realizations of disorder on a lattice with periodic boundary conditions. Solid line shows the ensemble average, while the shaded region indicates the region within one standard deviation of the ensemble average. (b) Similar, but for the ensemble averaged local Chern marker calculated in the lattice's center for $\rho = 10a_l$ at $E=0$. (c)-(f) Ensemble averaged variation of the local lattice properties with respect to restriction radius for five disorder strengths and 50 disorder configurations, local gap (c), Localizer gap (d), Local Chern marker (e), and $g_\rho/\mu_{\kappa,\rho}$ (f). The five disorder strengths are $\lambda = 0$, $\lambda = 1.5t$, 
    $\lambda = 2.0t$, $\lambda = 2.5t$, and $\lambda = 4.0t$. The strongest disorder strength is not shown in (f). Spectral localizer calculations use $\kappa = 0.2t/a_l$. The gray dashed line in (f) indicates $g_\rho/\mu_{\kappa,\rho} = 2$.}
    \label{fig:anderson}
\end{figure*}

In the work \cite{LSS} it was shown that in a random $p+ip$ superconductor in the intermediate coupling regime of $\lambda$, the spectral localizer is typically gapped even in the regime of Anderson localization where the Hamiltonian is not gapped. Using the new notion of local gap $g_\rho$ and Theorem~\ref{theo-localGap} one can understand this observation much better at least in the regime of small density of states (DOS). Indeed, while the norms $H_\omega(\lambda)(\imath \,\one +\frac{1}{\rho} D)^{-1}$ and $[D,H_\omega(\lambda)](\imath\,\one +\frac{1}{\rho} D)^{-1}$ do not fluctuate too much and agree with the norms without the resolvents, the crucial new ingredient for this mobility gap regime is that merely the local gap $g_\rho$ enters into the criterion \eqref{eq-kappaCond} in Theorem~\ref{theo-localGap}. To access it, the spectral analysis of $(H_\omega(\lambda)^2)_\rho$ should, according to the definition of the $\rho$-local gap, be carried out after restriction to $B_\rho(0)$ (Dirichlet boundary conditions), but if $\rho$ is sufficiently large most of the eigenvalues and eigenvectors are very close to the eigenvalues and eigenfunctions of the infinite volume operator $H_\omega(\lambda)^2$, except for those states localized at the boundary of $B_\rho(0)$. Hence one roughly has
$$
H_\omega(\lambda)^2
\;\approx\;
\sum_{j}(E_j)^2\,\big(|\phi_j\rangle\langle\phi_j|\big)_\rho
\;,
$$
where the $E_j$ are the random real eigenvalues of $H_\omega(\lambda)$ and $|\phi_j\rangle$ are the associated normalized eigenvector with localization center $c_j$. Due to the exponential localization of these states, the sum can effectively be cut to those $j$ with $c_j\in B_\rho(0)$. Because of the Poisson statistics mentioned above, one has (up to constant factors)
$$
\#
\big\{
E_j\,:\,c_j\in B_\rho(0)\,,\;E_j\in(-\delta,\delta)\big\}
\;\approx\;
\delta\,\rho^d\;\tfrac{d\Nn}{dE}(0)
\;,
$$
where $\frac{d\Nn}{dE}(0)$ is the density of states at energy $E=0$ (the choice of reference energy here). Then the expected value of the gap is roughly given by the smallest $\delta$ for which the r.h.s.\ is less than $1$, namely one infers
$$
\EM(g_\rho)
\;\approx\;
\frac{1}{\rho^d\;\frac{d\Nn}{dE}(0)}
\;.
$$
Numerics in Fig.~\ref{fig:anderson}(c) indeed show that average gap $\EM(g_\rho)$ decreases to $0$ as $\rho$ increases. Based on this expression for $\EM(g_\rho)$ and \eqref{eq-kappaCond3} (namely disregarding the improvements coming from the relative bounds on $H$ and its noncommutative derivative), one sees that a volume of size $\rho\leq\rho_c$  with roughly 
$$
\rho_c\;\approx\;
\Big(
C_F\,\|[H\|\,\|[D,\HP]\|\,\tfrac{d\Nn}{dE}(0)^2
\Big)^{-\frac{1}{2d-1}}
\;,
$$
allows to satisfy \eqref{eq-kappaCond3}, so that one can choose $\kappa\approx \frac{\EM(g_\rho)}{\rho}\approx \frac{1}{\rho^{d+1}}\,\frac{d\Nn}{dE}(0)^{-1}$. Theorem~\ref{theo-localGap} then implies that the spectral localizer $L_{\kappa,\rho}$ typically has an open gap and its local topological index can safely be computed. 

\vspace{.2cm}

However, the argument above is of interest only when the DOS is small ({\it e.g.} Lifshitz tail regime), because otherwise $\rho_c$ may be smaller than $1$. Numerics in Fig.~\ref{fig:anderson} show that the local topological index can actually be applied for much larger DOS. The above argument suggests that it is advantageous to work with relatively small $\rho$. Picking $\rho \approx 10 \,a_l$, one then infers from Fig.~\ref{fig:muGap}(a) that a good choice is $\kappa=0.2 (t/a_l)$ as then the maximal localizer gap is attained at $\rho=10 \,a_l$. Working with such a large $\kappa$ strengthens the locality property of the spectral localizer. Fig.~\ref{fig:anderson}(d) then shows that the localizer gap is still large, even for relatively large $\lambda$, and this even though the local gap of the Hamiltonian is relatively small so that this regime is not adequately covered by the above reasoning based on Theorem~\ref{theo-localGap}. Most interesting are the plots for $\lambda=2.5t$. For $\lambda=2t$, the bulk gap closes and for $\lambda=2.5t$, Fig.~\ref{fig:anderson}(a) clearly shows that the DOS is positive. Nevertheless, Fig.~\ref{fig:anderson}(c) indicates that the local topological index shows no fluctuations yet {and that the local gap remains open indicating that the local Chern number is stable}. For larger $\lambda$, however, the average localizer gap is smaller and has larger fluctuations, and the same holds for the local topological index, {indicating the instability of this marker}. As $\rho\to\infty$ and $\kappa\to0$, we expect the topological transition to appear at the single critical value $\lambda_c$. 

%\vspace{.2cm}

%%%%%%%%%%%%%%%%%%%%%%%%%%%%%%%%%%%%%%%%%%%%%
\section{Stability of the spectral flow of the spectral localizer}
\label{sec-SFstab}

Let $H$ have $\rho$-local spectral gaps at $x_0$ and at $x_1$ and suppose that both local gaps satisfy \eqref{eq-kappaCond}. Then by Theorem~\ref{theo-localGap} there exists a $\kappa$ such that $\mu_{\kappa,\rho}(H,x_0)>0$ and $\mu_{\kappa,\rho}(H,x_1)>0$. Otherwise stated, the two spectral localizers $L_{\kappa,\rho}(H,x_0)$ and $L_{\kappa,\rho}(H,x_1)$ are invertible. Recall ({\it e.g.}\ \cite{DSW}) that the spectral flow of a continuous path of selfadjoint matrices with invertible endpoints is defined as the sum of all eigenvalue crossings through $0$ weighted by the sign of the speed, and it is also equal to the difference of half-signatures of the endpoints. One may now be tempted to consider the straight-line path $t\in[0,1]\mapsto x_t=x_0+t(x_1-x_0)$ and then the spectral flow of $t\in[0,1]\mapsto L_{\kappa,\rho}(H,x_t)$ connecting $L_ {\kappa,\rho}(H,x_0)$ and $L_{\kappa,\rho}(H,x_1)$, but this path does not consist of matrices of the same size. One way to circumvent this difficulty is to choose a subset $\Lambda\subset \ZM^d$ that covers $B_\rho(x_t)\cap\ZM^d$ for all $t\in[0,1]$. Then $t\in[0,1]\mapsto L^\Lambda_{\kappa}(H,x_t)$ is a path of selfadjoint matrices of same size. Due to Proposition~\ref{prop-IncreasingVolume}, the signatures of its endpoints are known and equal to the (possibly different) local topological indices. Therefore now the spectral flow is well-defined and satisfies
\begin{equation}
\label{eq-SFInd}
\SF\big(t\in[0,1]\mapsto L^\Lambda_{\kappa}(H,x_t)\big)
\;=\;
\LocInd_{\kappa,\rho}(H,x_1)\,-\,\LocInd_{\kappa,\rho}(H,x_0)
\;.
\end{equation}
Actually, replacing the straight-line path by any other path in $\Lambda$ leads to the same formula. Let us stress that along the path the local gap condition is not necessarily satisfied. Indeed, when crossing topological phase boundaries, it will not be so and this is the situation where \eqref{eq-SFInd} is of interest. The spectral flow has again stability properties:

%%%%%%%%%%%%%%%%%%%%%%%%
\begin{proposition}
\label{prop-SFStability}
Suppose that a weakly local Hamiltonian $H$ has a $\rho$-local gaps $g_\rho(H,x_0)$ and  $g_\rho(H,x_1)$ at $x_0$ and $x_1$ respectively, which both satisfies \eqref{eq-kappaCond}. Then for any weakly local perturbation $W$ satisfying $\supp(W)\cap B_\rho(x_0)=\emptyset$ and $\supp(W)\cap B_\rho(x_1)=\emptyset$, one has
$$
\SF\big(t\in[0,1]\mapsto L^\Lambda_{\kappa}(H+W,x_t)\big)
\;=\;
\SF\big(t\in[0,1]\mapsto L^\Lambda_{\kappa}(H,x_t)\big)
\;.
$$
\end{proposition}
%%%%%%%%%%%%%%%%%%%%%%%%

\noindent {\bf Proof.} The claim follows from the homotopy invariance of the spectral flow applied to the path (of paths) $s\in[0,1]\mapsto L^\Lambda_{\kappa}(H+s W,x_t)$, because the invertibility at both the left and right endpoint is guaranteed by  Proposition~\ref{prop-IncreasingVolume}, namely  $s\in[0,1]\mapsto L^\Lambda_{\kappa}(H+s W,x_0)$ and $s\in[0,1]\mapsto L^\Lambda_{\kappa}(H+s W,x_1)$ are both paths in the invertible matrices.
\hfill $\Box$

\vspace{.2cm}

\begin{figure*}[t]
    \centering
    \includegraphics{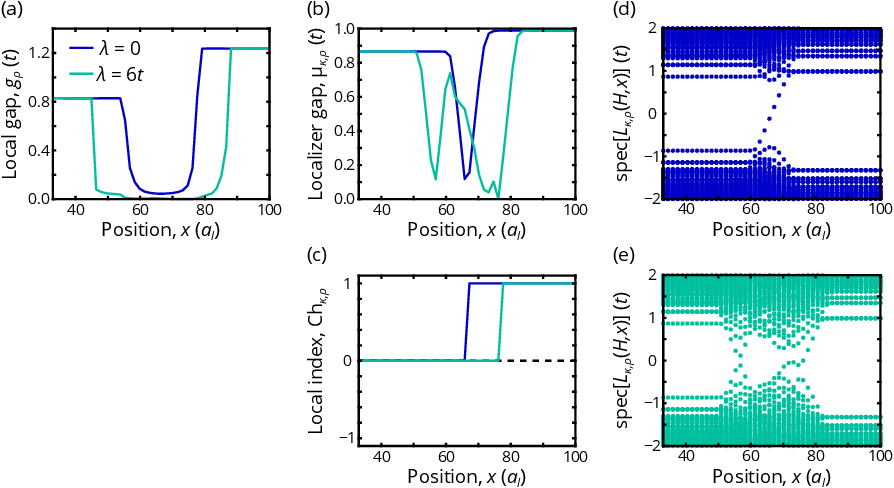}
    \caption{Simulations demonstrating the stability of the spectral localizer's spectral flow in the presence of disorder. The local properties of a $90 \times 90$ site heterostructure consisting of half massive graphene and half Haldane lattice are compared against a disordered variant of the system as the probe position $x$ is varried across the system. The clean system uses the same parameters and orientation as Fig.~\ref{fig:gapDemo}c, while the added disorder is on-site with $\lambda = 6t$, see Eq.~\eqref{eq:Hdisorder}, but only in a disk of radius $10a_l$ from the lattice's center (i.e., the disorder intersects the heterostructure's material interface). (a)-(e) Specifically, the local gap (a), localizer gap (b), local index (c), and low-lying spectrum of the spectral localizer for the clean (d) and disordered (e) systems are shown as the evaluation position is moved across the lattice's middle. Simulations use $\rho = 12a_l$ and $\kappa = 0.2t/a_l$.}
    \label{fig:specFlow}
\end{figure*}

Let us stress that Proposition~\ref{prop-SFStability} does not exclude that $\supp(W)$ has non-trivial intersection with the path from $x_0$ to $x_1$. If this is the case, the perturbations may create new zero eigenvalues of the spectral localizer, that is points $x_t$ at which the stability gap $\mu^\Lambda_{\kappa}(H,x)$ vanishes, but according to Proposition~\ref{prop-SFStability} this will not alter the spectral flow. Let us also point out that Proposition~\ref{prop-SFStability} does not use the improved criterion \eqref{eq-kappaCond}, and could also be stated only based on \eqref{eq-kappaCond3}, which involves local gaps though, other than the global gaps used in prior works. These results are illustrated in Fig.~\ref{fig:specFlow}, which compares the local gap, localizer gap, and spectral flow of clean and perturbed heterostructrues in which the defect is limited in its spatial extent such that each constituent material still exhibits a clean bulk region where $\mu_{\kappa,\rho}(H,x_0)>0$ and $\mu_{\kappa,\rho}(H,x_1)>0$. Even though the straight-line path between $x_0$ and $x_1$ intersects the strongly perturbed region, where both the local gap and localizer gap close, the spectral flow along the full path is preserved.

%%%%%%%%%%%%%%%%%%%%%%%%%%%%%%%%%%%%%%%%%%%%%
\section{Bounds on the eigenvalue slope of the localizer}
\label{sec-SFstab2}

The set-up for this section is the same as in Section~\ref{sec-SFstab}, namely we consider the spectral flow of a path $t\in[0,1]\mapsto L_{\kappa,\rho}(H,x_t)$ of spectral localizers along a physical path $t\in[0,1]\mapsto x_t=x_0+t(x_1-x_0)$ which crosses a topological interface. Proposition~\ref{prop-SFStability} already shows that this spectral flow is constant under a perturbation satisfying weak assumptions which allow the perturbation even to have a support intersection with the trace of the physical path. However, such a perturbation may change the low-lying spectrum of the spectral localizer along the path considerably and can potentially lead to several new eigenvalue crossings of the spectral localizer, see Fig.~\ref{fig:specFlow}. The aim of this section is to provide a criterion on $W$, again involving the geometric separation of its support to the path, that ensures that the transversality of the eigenvalue crossings is conserved. Based on Theorem~\ref{theo-localGap}, such a result is to be expected, but does not follow directly from Theorem~\ref{theo-localGap}. Here such a stability result for the transversality of eigenvalue crossings of the spectral localizer will be achieved by second order perturbation theory. However, the higher order terms are only controlled in powers of the operator norm of $W$ (unlike in Theorem~\ref{theo-localGap}). 

\vspace{.2cm}

Let us now describe the more concrete set-up. Suppose that there is a $t_c$ such that $\mu^\Lambda_{\kappa}(H,x_{t_c})=0$ and that the kernel of $L^\Lambda_ {\kappa}(H,x_{t_c})$ is one-dimensional. Hence there is a simple eigenvalue $\mu(x)$ of $L^\Lambda_ {\kappa}(H,x)$ which crosses $0$ at $x_{t_c}$. Finally let us also consider that the eigenvalue slope $(\partial_x\mu)(x_{t_c})=(\partial_t\mu)(x_{t})|_{t=t_c}$ at the crossing is bounded away from $0$, namely the crossing is transversal. The aim of this section is to provide a quantitative bound on a perturbation $s W$ which assures that the eigenvalue $\mu(x,s)$ of $L^\Lambda_ {\kappa}(H+s W,x)$ still has a transversal eigenvalue crossing. Clearly the value of $x(s)$ of the crossing $\mu(x(s),s)=0$ will typically be shifted away from $x_{t_c}$, so that the stability result is rather on the eigenvalue slope. This will be attained per a two-parameter perturbation theory as outlined in Appendix~\ref{app-TwoPar}. For the statement of the result, let us {recall that $|X-x|=(\sum_{j=1}^d(X_j-x_j)^2)^{\frac{1}{2}}$ is the operator} measuring the Euclidean distance from $x\in\RM^d$.

%%%%%%%%%%%%%%%%%%%%%%%%
\begin{proposition}
\label{prop-SlopeStability}
For the set-up described above and with the notation $\langle X\rangle_x=\one+|X-x|$,
$$
\Big|\mu(x_{t_c},s)
\,-\,
\mu(x_{t_c},0)
\Big|
\;\leq\;
s\,\big\|\langle X\rangle_{x_{t_c}}^{-1} W \langle X\rangle_{x_{t_c}}^{-1}\big\|\;
\big(1+\tfrac{1}{\kappa}\|H\|\big)^2 (1+\Oo(s))
$$
and
$$
\Big|(\partial_x\mu)(x_{t_c},s)
\,-\,
(\partial_x\mu)(x_{t_c},0)
\Big|
\;\leq\;
s\,
\tfrac{d}{g_L} \,
\|W \langle X\rangle_{x_{t_c}}^{-1}\big\|\;
\big(\kappa+\|H\|\big)(1+\Oo(s))
\;,
$$
where $g_L$ is the distance from $\mu(x_{t_c},0)=0$ to the rest of the spectrum of $L^\Lambda_\kappa(x_{t_c},0)$, and $\Oo(s)$ means, as usual, that the error is uniformly bounded by $Cs$ for some constant $C$.
\end{proposition}
%%%%%%%%%%%%%%%%%%%%%%%%

The main insight of Proposition~\ref{prop-SlopeStability} is that the low lying eigenvalue of $L^\Lambda_ {\kappa}(H,x)$ is not substantially altered if a perturbation $s W$ is added to the Hamiltonian that is localized far from the locality center $x$. Indeed, the operators $\langle X\rangle_x^{-1} W \langle X\rangle_x^{-1}$ and $W \langle X\rangle_x^{-1}$ are then considerably smaller in norm {than} the norm of $W$ itself. As is shown in the proof, this is a result of the fact that the eigenvector of the lowest eigenvalue of $L^\Lambda_ {\kappa}(H,x)$ has rapid drop-off from $x$ (more precisely, the eigenvector has a second moment). 

\vspace{.2cm}

The second insight is that if the slope $(\partial_x\mu)(x_{t_c},0)$ at the eigenvalue crossing is non-vanishing and the perturbation is sufficiently small or distant in the quantitative manner stated in Proposition~\ref{prop-SlopeStability}, the slope does not change sign. Combined with the first estimate in Proposition~\ref{prop-SlopeStability}, this shows that the transversality of the crossing as well as its contribution to the spectral flow is not altered. Hence when the argument is applied to a path which has only transversal eigenvalue crossings through $0$, one deduces again the stability of the spectral flow proved in Proposition~\ref{prop-SFStability}, albeit only for weak perturbations. 

\vspace{.2cm}

The first auxiliary result for the proof of Proposition~\ref{prop-SlopeStability} is that any eigenvector of the spectral localizer has a strong locality property. For later use, the us formulate the statement directly for an arbitrary subset $\Lambda\subset\ZM^d$ with associated  surjective partial isometry  $\pi_\Lambda:\ell^2(\ZM^d,\CM^L)\to\ell^2(\Lambda,\CM^L)$. Then for an operator $A$ on $\ell^2(\ZM^d,\CM^L)$ its restriction to $\Lambda$ is $A^\Lambda=\pi_\Lambda A\pi_\Lambda^*$. {All these notations are naturally extended to the tensor product $\ell^2(\ZM^d,\CM^{L})\otimes\CM^{d'}\cong\ell^2(\ZM^d,\CM^{Ld'})$ with the Clifford degrees of freedom.} The spectral localizer on $\Lambda$ is then denoted by $L^\Lambda_{\kappa}(H,x)=L_{\kappa}(H,x)^\Lambda$.

%%%%%%%%%%%%%%%%%%%%%%%%
\begin{lemma}
\label{lem-KernelVecEst}
Let $\Lambda\subset\ZM^d$. A normalized eigenvector {$\phi\in\ell^2(\Lambda,\CM^{2Ld'})$} of $L^\Lambda_\kappa(H,x)$ for the eigenvalue $\mu\in\RM$ satisfies
$$
\||X-x|\phi\|
\;\leq\;
\tfrac{1}{\kappa}(\|H\|+|\mu|)
\;.
$$
\end{lemma}
%%%%%%%%%%%%%%%%%%%%%%%%

\noindent {\bf Proof.} 
Decompose $\phi=\binom{\phi_+}{\phi_-}$ into upper and lower components {$\phi_\pm\in\ell^2(\Lambda,\CM^{L\frac{d'}{2}})$} in the grading \eqref{eq-SpecLocDef} of the spectral localizer. Then
$$
\mu\binom{\phi_+}{\phi_-}
\;=\;
L^\Lambda_\kappa(H,x)\phi
\;=\;
\pi^\Lambda
\binom{-H(\pi^\Lambda)^*\phi_++\kappa D_0(x)^*(\pi^\Lambda)^*\phi_-}{
\kappa D_0(x)(\pi^\Lambda)^*\phi_++ H(\pi^\Lambda)^*\phi_-}
\;=\;
\binom{-H^\Lambda \phi_++\kappa D^\Lambda_0(x)^*\phi_-}{
\kappa D_0^\Lambda(x)\phi_++ H^\Lambda\phi_-}
\;,
$$
where as above  $D^\Lambda_0(x)=\pi^\Lambda D_0(x)(\pi^\Lambda)^*$ and  $H^\Lambda=\pi^\Lambda H(\pi^\Lambda)^*$. Hence
$$
\|\kappa D^\Lambda_0(x)^*\phi_-\|\;=\;\|(H^\Lambda+\mu\one_\Lambda)\phi_+\|
\;,
\qquad
\|\kappa D^\Lambda_0(x)\phi_+\|\;=\;\|(H^\Lambda-\mu\one_\Lambda)\phi_-\|
\;.
$$
Now $|D^\Lambda_0(x)|=|(D^\Lambda_0(x))^*|=|X-x|$ so that
\begin{align*}
\|\,|X-x|\phi\|^2
&
\;=\;
\| D^\Lambda_0(x)^*\phi_-\|^2
+
\| D^\Lambda_0(x)\phi_+\|^2
\\
&
\;\leq\;
\tfrac{1}{\kappa^2}\big(\|H-\mu\one\|^2\|\phi_-\|^2+\|H+\mu\one\|^2\|\phi_+\|^2\big)
\\
&
\;\leq\;
\tfrac{1}{\kappa^2}(\|H\|+|\mu|)^2(\|\phi_-\|^2+\|\phi_+\|^2)
\;,
\end{align*}
implying the claim.
\hfill $\Box$

\vspace{.2cm}

\noindent {\bf Proof} of Proposition~\ref{prop-SlopeStability}. 
One may assume that $x_{t_c}=0$ after shifting the positions. Recall that the simple eigenvalue crossing at $0$ is denoted by $\mu(x,s)$, namely $\mu(0,0)=0$. Let $\phi$ be as in Lemma~\ref{lem-KernelVecEst}. Then first order perturbation theory, see Appendix~\ref{app-TwoPar}, implies
$$
\mu(0,s)
\;=\;
\langle \phi|s \,\diag(-W,W) |\phi\rangle +\Oo(s^2)
\;=\;
s
\big(\langle \phi_-| W |\phi_-\rangle-\langle \phi_-| W |\phi_-\rangle\big) +\Oo(s^2)
\;.
$$
On the other hand, using the bound of (the proof of) Lemma~\ref{lem-KernelVecEst},
\begin{align*}
\langle \phi_\pm| W |\phi_\pm\rangle
&
\;=\;
\langle \phi_\pm|\langle X\rangle_0\langle X\rangle_0^{-1} W \langle X\rangle_0^{-1}\langle X\rangle_0|\phi_\pm\rangle
\\
&
\;\leq\;
\big\|\langle X\rangle_0^{-1} W \langle X\rangle_0^{-1}\big\|\;
\|\langle X\rangle_0|\phi_\pm\rangle\|^2
\\
&
\;\leq\;
\big\|\langle X\rangle_0^{-1} W \langle X\rangle_0^{-1}\big\|\;
\big(1+\tfrac{1}{\kappa}\|H\|\big)^2\|\phi_\pm\|^2
\;,
\end{align*}
which implies the first claim. As to the second, let us note that the smallest eigenvalue $\mu(x,s)$ is real analytic in both parameters (see Appendix~\ref{app-TwoPar}) and can (recall $x_{t_c}=0$ here) hence be expended as $\mu(x,s)= \mu_{1,0}x+\mu_{0,1}s+\mu_{1,1}xs+\Oo(x^2,s^2)$. Now $\mu_{1,0}=(\partial_x\mu)(0,0)$ and the difference $(\partial_x\mu)(0,s)-(\partial_x\mu)(0,0)$ is hence equal to $\mu_{1,1}s+\Oo(s^2)$ so that the claim is a statement about the coefficient $\mu_{1,1}$. This is given by the perturbative formula \eqref{eq-SecondOrderPert} where here $V=\partial_x L^\Lambda_\kappa(H,x)$. Due to the definition \eqref{eq-SpecLocDef}, this is in turn given by
$$
V
\;=\;
\kappa\,\sum_{j=1}^{d}\gamma_j
\;,
$$
so that $\|V\|\leq \kappa d$. Furthermore, here $P_{0,0}=|\phi\rangle\langle\phi|$ and $L=L^\Lambda_\kappa(0,0)$ so that one gets the norm estimate $\|(\one-P_{0,0})L^{-1}(\one-P_{0,0})\|\leq \frac{1}{g_L}$. Replacing this in  \eqref{eq-SecondOrderPert} one deduces
$$
|\mu_{1,1}|
\;\leq\;
\kappa \,d\,\tfrac{1}{g_L} \|W\phi\|
\;\leq\;
\kappa \,d\,\tfrac{1}{g_L} \,
\|W \langle X\rangle_0^{-1}\big\|\;
\big(1+\tfrac{1}{\kappa}\|H\|\big)
\;.
$$
This directly shows the desired bound.
\hfill $\Box$

\appendix

%%%%%%%%%%%%%%%%%%%%%%%%%%%%%%%%%%%%%%%%%%
\section{The tapering estimate}
\label{app-tapering}

A tapering function is an even differentiable, but non-analytic function $F:\RM\to [0,1]$ with the property that $F([-\frac{1}{2},\frac{1}{2}])=1$ and $F(\RM\setminus (-1,1))=0$. Associated to $F$ and $\rho>0$, one then sets $F_\rho(x)=F(\frac{x}{\rho})$. Let us first show how to derive the tapering estimate \eqref{eq-TaperingEst} (by slightly extending the argument in \cite{BR}). First of all, recall that for any differentiable function $F$, the derivative of which has an integrable Fourier transform,
\begin{align*}
F(x)
\,=\,F(0)+\int^x_0dy \,F'(y)
% \\ & 
\,=\,F(0)+\int^x_0dy\int dp \,e^{\imath py}\,\widehat{F'}(p)
%\\
%& \;=\;f(0)\,+\,\int dp\,g(p) \,\int^x_0dy\,e^{\imath py}
%\\ & 
\,=\,F(0)+\int dp\, \widehat{F'}(p) \,\frac{e^{\imath px}-1}{\imath p}
\;,
\end{align*}
where the Fourier transform of a function $G$ is here defined by $\widehat{G}(p)=\frac{1}{2\pi}\int_{-\infty}^\infty e^{-\imath p x}G(x) dx$. Hence by functional calculus of $D$ and setting $R=\imath\,\one+\frac{1}{\delta}D$, one has the identity
$$
[F_\rho(D),\HP]R^{-1}
\;=\;
\int dp \,\frac{\widehat{F'_{\rho}}(p)}{\imath p}\,[e^{\imath pD},\HP]R^{-1}
\;.
$$
As $[D,R]=0$, DuHamel's formula {$[e^{\imath pD},\HP]=\int^1_0ds\;e^{\imath (1-s)pD}[\imath pD,\HP]e^{\imath spD}$} implies 
\begin{equation}
\label{eq-DuhamelCommutator}
[F_\rho(D),\HP]R^{-1}
\;=\;
\int dp \,\frac{\widehat{F'_{\rho}}(p)}{\imath p}\,\int^1_0ds\;e^{\imath (1-s)pD}[\imath pD,\HP]R^{-1}e^{\imath spD}
\;.
\end{equation}
%
% $\|[e^{\imath pD},\HP]R^{-1}\|\leq |p|\,\|[D,\HP]R^{-1}\|$. 
Therefore a norm estimate allows to conclude that
$$
\|[F_\rho(D),\HP]R^{-1}\|
\;\leq\;
\|\widehat{F'_{\rho}}\|_1\,
\|[D,\HP]R^{-1}\|
\;=\;
\tfrac{1}{\rho}\,\|\widehat{F'}\|_1\,
\|[D,\HP]R^{-1}\|
\;,
$$
This shows that \eqref{eq-TaperingEst} holds for a constant $C_F=\|\widehat{F'}\|_1$. The references \cite{LS2,DSW} construct a tapering function $F$ explicitly and show $\|\widehat{F'}\|_1=8$, namely $C_F\leq 8$ for this function. 

\vspace{.2cm}

The following argument improves this bound. More precisely, another function $F$ is constructed for which it is shown that $C_F\leq 4.56$. The construction of $F$ will start from an arbitrary integrable function $\phi:[0,1]\mapsto \RM_{\geq}=[0,\infty)$. Then let us introduce a step function $\varphi:\RM\to [0,1]$ by 
\begin{align*}
\varphi(x)
&
\;=\;
\left\{
\begin{array}{cc}
0\;, & x<0\;,
\\
\tfrac{1}{C_\phi}\,\int^x_0dy\,\phi(y) \;,
& x\in[0,1]\;,
\\
1\;, & x>1
\;,
\end{array}
\right.
\qquad
\mbox{\rm where }\;\;
C_\phi\;=\;
\int^1_0dy\,\phi(y)
\;.
\end{align*}
The function $F$ is then built from two such step functions:
\begin{equation}
\label{eq-Fbuild}
F(x)
\;=\;
\varphi(2x+2)
-
\varphi(2x-1)
\;.
\end{equation}
The Fourier transform of the derivative $F'$ can now be computed:
\begin{align*}
\widehat{F'}(p)
&
\;=\;
\frac{1}{2\pi}\int_{-\infty}^\infty e^{-\imath p x}F'(x)\, dx
\\
&
\;=\;
\frac{1}{2\pi}\int_{-\infty}^\infty e^{-\imath p x}\big(2\,\varphi'(2x+2)-2\,\varphi'(2x+1)\big) dx
\\
&
\;=\;
\frac{1}{2\pi}\int_{0}^1 e^{-\imath p \frac{y}{2}}\,\varphi'(y)\,e^{-\imath p }dy
\,-\,
\frac{1}{2\pi}\int_{0}^1 e^{-\imath p \frac{y}{2}}\,\varphi'(y)\,e^{-\imath \frac{p}{2} }dy
\\
&
\;=\;
\frac{1}{2\pi\,C_\phi}\int_{0}^1 e^{-\imath p \frac{y}{2}}\,\phi(y)\,e^{-\imath p }dy
\,-\,
\frac{1}{2\pi\,C_\phi}\int_{0}^1 e^{-\imath p \frac{y}{2}}\,\phi(y)\,e^{-\imath \frac{p}{2} }dy
\\
&
\;=\;
\frac{1}{C_\phi}\,\widehat{\phi}(\tfrac{p}{2})\,\big(e^{-\imath p }
\,-\,e^{-\imath \frac{p}{2} }\big)
\\
&
\;=\;
\frac{2}{\imath\,C_\phi}\,\,\widehat{\phi}(\tfrac{p}{2})\,
\sin(\tfrac{p}{4})
\,e^{-\imath \frac{3p}{4} }
\;.
\end{align*}
Hence
\begin{align}
\|\widehat{F'}\|_{L^1}
&
\;=\;
\frac{2}{C_\phi}
\,
\|\widehat{\phi}(\tfrac{\cdot}{2})\sin(\tfrac{\cdot}{4})\|_{L^1}
%\nonumber
%\\
%&
\;=\;
\frac{4}{C_\phi}
\int_{-\infty}^\infty |\widehat{\phi}(p)|\,
|\sin(\tfrac{p}{2})|\,dp
%\nonumber
%\\
%&
%\;=\;
%\frac{4}{C_\phi}
%\int_{-\infty}^\infty \Big|\frac{1}{2\pi}\int^1_0 e^{-\imath p x}\,\phi(x)\,dx\Big|\,
%|\sin(\tfrac{p}{2})|\,dp
\;.
\label{eq-FPrimeHat}
\end{align}
Up to now, no approximations have been made. One can now choose the function $\phi$ and attempt to minimize the r.h.s.\ of \eqref{eq-FPrimeHat} over these choices. For this purpose, it is appealing to take functions $\phi$ which vanish at the boundaries $0$ and $1$. Hence, let us consider the family of functions $\phi_k$, depending on $k\geq 0$, given by
$$
\phi_k(x)
\;=\;
x^k(1-x)^k
\;.
$$
For integer $k\in\NM_0$, the associated normalization constants can be computed explicitly to be
$$
C_{\phi_k}\;=\;
\int^1_0dy\,y^k(1-y)^k
\;=\;
\frac{(k!)^2}{(2k+1)!}
\;,
$$
and $\varphi_k$ is the regularized incomplete Beta function:
$$
\varphi_k(x)\;=\;I_x(k+1,k+1)
\;,
\qquad
x\in [0,1]
\;.
$$
Explicitly, still for $x\in [0,1]$, one has $\varphi_1(x)=3x^2-2x^3$. 
%and more generally:
%
%$$
%\varphi_k(x)
%\;=\;
%\frac{(2k+1)!}{(k!)^2}
%\int_0^x y^k (1 - y)^k \, dy 
%\;=\; 
%\frac{(2k+1)!}{(k!)^2}
%\sum_{j=0}^k (-1)^j \binom{k}{j} \,\frac{x^{k + j + 1}}{k + j + 1}
%\;.
%$$
%
Next the Fourier transform of $\phi_k$ is needed:
\begin{align*}
\widehat{\phi}_k(p)
%&
%\;=\;
%\frac{1}{2\pi}\int_{-\infty}^\infty e^{-\imath p x}\phi_k(x)\, dx
%\\
&
\;=\;
\frac{1}{2\pi}
\,
\int_{0}^1 e^{-\imath p x}
\,x^k(1-x)^k
\, dx
%\\
%&
%\;=\;
%\frac{1}{2\pi} \,
% \frac{(k!)^2}{(2k+1)!} \;_1F_1(k+1,2k+2,-\imath p)
%\\
%&
\;=\;
\frac{C_{\phi_k}}{2\pi}
\;
_1F_1(k+1,2k+2,-\imath p)
\;,
\end{align*}
where $_1F_1$ is the confluent hypergeometric function of the first kind (also called Kummer's function). Hence replacing leads to tapering functions $F_k$ which, according to \eqref{eq-FPrimeHat}, satisfy
$$
\|\widehat{F'_k}\|_{L^1}
\;=\;
\frac{2}{\pi} \int_{-\infty}^\infty | _1F_1(k+1,2k+2,-\imath p)|\,
|\sin(\tfrac{p}{2})|\,dp
\;.
$$
Mathematica then gives:
$$
\|\widehat{F_0'}\|_{L^1}
\,\approx\,
9.16
\;,
\qquad
\|\widehat{F_1'}\|_{L^1}
\,\approx\,
4.56
\;,
\qquad
\|\widehat{F_2'}\|_{L^1}
\,\approx\,
5.12
\;,
\qquad
\|\widehat{F_3'}\|_{L^1}
\,\approx\,
5.75
%\;,
%\quad
%\|\widehat{F_4'}\|_{L^1}
%\,\approx\,7.28
\;.
$$
Intuitively, these findings are explained as follows. As $k$ increases, the regularity of the functions $\phi_k$ increases (namely, $\phi_k$ is $C^{k-1}$). This leads to a faster decay of the Fourier transform as $k$ increases. On the other hand, the central peak of the Fourier transform also becomes larger and thus gives a larger contribution to the $L^1$-norm. Hence there is a trade-off between these two effects. Numerics show that within the family $\phi_k$ for positive real $k$, the minimal value of $\|\widehat{F'_k}\|_{L^1}$ is actually attained at $k=1$. Hence let us compute the Fourier transform of $\phi_1$ more explicitly. 
\begin{align*}
\widehat{\phi}_1(p)
&
%\;=\;
%\frac{1}{2\pi}\int_{-\infty}^\infty e^{-\imath p x}\phi_1(x)\, dx
%\\
%&
\;=\;
\frac{1}{2\pi}\int_{0}^1 e^{-\imath p x}(x-x^2)\, dx
%\\
%&
\;=\;
\frac{1}{2\pi}
\,\frac{1}{p^3}
\big(
4\,\sin(\tfrac{p}{2})-2\,p\,\cos(\tfrac{p}{2})\big)\,e^{-\imath \frac{p}{2}}
\;.
\end{align*}
Note that in spite of the factor $\frac{1}{p^3}$ there is no singularity at $p=0$, as a Taylor expansion of the other factor readily shows. As $C_1=\frac{1}{6}$, one hence gets from \eqref{eq-FPrimeHat} that
\begin{align*}
\|\widehat{F'_1}\|_{L^1}
\;=\;
\frac{12}{\pi}
\int_{-\infty}^\infty  \,\frac{1}{|p|^3}
\big|
4\,\sin(\tfrac{p}{2})-2\,p\,\cos(\tfrac{p}{2})\big|\,
|\sin(\tfrac{p}{2})|\,dp
\;,
\end{align*}
and evaluating this integral gives the value $4.56$ stated above. We attempted several other natural choices for the function $\phi$, but none produced a smaller value $\|\widehat{F'}\|_{L^1}$.

%
%It is a somewhat surprising fact that this leads to very similar values $C_F$.

\vspace{.2cm}

This does not exclude the possibility to improve the bound \eqref{eq-TaperingEst} though. Indeed, when taking the operator norm of \eqref{eq-DuhamelCommutator} one disregards oscillations stemming from the exponential factors. Let us first support this claim by an explicit study of the bound for the one-dimensional discrete Schr\"odinger operator $H=S+S^*+V(X)$ on $\ell^2(\ZM)$ where $S$ is the shift and $V$ is a real valued function.  For the computation of the commutator $[F_\rho(X),H]$ with $F$ given by \eqref{eq-Fbuild}, let us first note that the two summands of $F$ are spatially separated and hence it is sufficient to compute the norm of one of them. Moreover, the function can be shifted, provided $\rho$ is an integer,
$$
\|[F_\rho(X),H]\|
\;=\;
\|[\varphi(\tfrac{2}{\rho}X+2),H]\|
\;=\;
\|S^\rho[\varphi(\tfrac{2}{\rho}X),H](S^\rho)^*\|
\;=\;
\|[\varphi(\tfrac{2}{\rho}X),H]\|
\;.
$$
Therefore focus is on estimating the latter norm. This will be carried out for $k=1$, namely $\varphi_1(x)=3x^2-2x^3$ for $x\in[0,1]$. Because $H$ has a hopping range $1$, the commutator $[\varphi(\tfrac{2}{\rho}X),H]$ is supported on $\ZM\cap [-1,\frac{\rho}{2}+1]$. On the boundary sites, the norm is small though so that one can essentially restrict to the central part:
$$
\big\|[\varphi_1(\tfrac{2}{\rho}X),H]\big\|
\;=\;
\big\|[\tfrac{12}{\rho^2}(X^2-\tfrac{1}{\rho}\,\tfrac{4}{3}\,X^3),S^*+S]|_{[0,\frac{\rho}{2}]}\big\|
\,+\,
\Oo(\tfrac{1}{\rho^2})\;,
$$
where the restriction is strictly speaking on the sites $\ZM\cap [0,\frac{\rho}{2}]$ and the norm difference is bounded by $\frac{12}{\rho^2}$. Next one can use the commutation relation $[X,S]=S$ and separately bound the summands with $S$ and those with $S^*$, which leads to a supplementary factor $2$:
\begin{align*}
\big\|[F_\rho(X),H]\big\|
&
\;\leq\;
\tfrac{12}{\rho^2} \,2
\big\|
XS+SX-\tfrac{1}{\rho}\,\tfrac{4}{3}(SX^2+XSX+X^2S)|_{[0,\frac{\rho}{2}]}\big\|
\,+\,
\Oo(\tfrac{1}{\rho^2})
\\
&
\;=\;
\tfrac{24}{\rho^2} 
\big\|
\big( 2X-1 -\tfrac{4}{\rho}(X^2-X)\big)S |_{[0,\frac{\rho}{2}]}\big\|
\,+\,
\Oo(\tfrac{1}{\rho^2})
\\
&
\;=\;
\tfrac{24}{\rho^2} 
\big\|
2X-1 -\tfrac{4}{\rho}(X^2-X)|_{[0,\frac{\rho}{2}]}\big\|
\,+\,
\Oo(\tfrac{1}{\rho^2})
\\
&
\;\leq\;
\tfrac{24}{\rho^2} \,\tfrac{\rho}{4}
\,+\,
\Oo(\tfrac{1}{\rho^2})
\\
&
\;\leq\;
\tfrac{6}{\rho} 
\,+\,
\Oo(\tfrac{1}{\rho^2})
\\
&
\;=\;
\tfrac{3}{\rho}\,\big\|[X,H]\big\| 
\,+\,
\Oo(\tfrac{1}{\rho^2})
\;,
\end{align*}
where in the last step it was used that $[X,H]=S-S^*$ has operator norm $2$. One can also include the resolvent of $X$. As $\|(S-S^*)(\imath\one +\frac{1}{\delta}X)^{-1}\|=2-\Oo(\frac{1}{\delta})$, one concludes
$$
\big\|[F_\rho(X),H](\imath\one +\tfrac{1}{\delta}X)^{-1}\big\|
\;\leq\;
\tfrac{3}{\rho}\,\big\|[X,H](\imath\one +\tfrac{1}{\delta}X)^{-1}\big\| 
\,+\,
\Oo(\tfrac{1}{\rho^2},\tfrac{1}{\delta \rho})
\;.
$$
Hence $C_F\leq 3$, up to the lower order summands. This shows that the norm estimate of \eqref{eq-DuhamelCommutator} is not optimal in this special case a Hamiltonian with hopping range $1$, because for $\varphi_1$ it only leads to $C_F\approx 4.56$. For operators with a larger range, it is possible, but tedious to generalize the above estimate (this leads to a larger constant $C_F$ though). One can also carry out the calculation for $\varphi_k$ with higher integer $k$. Instead of providing all the details, let us rather state the outcome, namely for $F$ constructed from $\phi_k$ one obtains
$$
\big\|[F_\rho(X),H]\big\|
\;\leq\;
\tfrac{1}{\rho}\,2\,(C_{\phi_k})^{-1}\,\phi_k(\tfrac{1}{2})\,
\big\|[X,H]\big\| 
\,+\,
\Oo(\tfrac{1}{\rho^2})
\;=\;
\tfrac{1}{\rho}\,
\tfrac{2\,(2k+1)!}{(k!)^22^{2k}}\,\big\|[X,H]\big\| 
\,+\,
\Oo(\tfrac{1}{\rho^2})
\;.
$$
Based on this, one readily checks again that $k=1$ is the optimal choice leading to the tightest bound.  The argument above directly transposes to the Su-Schrieffer-Heeger (SSH) model \cite{SSH} if matrix degrees of freedom are included. It also allows to deal with a higher dimensional Laplacian, by dealing with each dimension separately and then using $\|[D,\HP]\|=2\,d$. 

\begin{figure*}[t]
    \centering
    \includegraphics{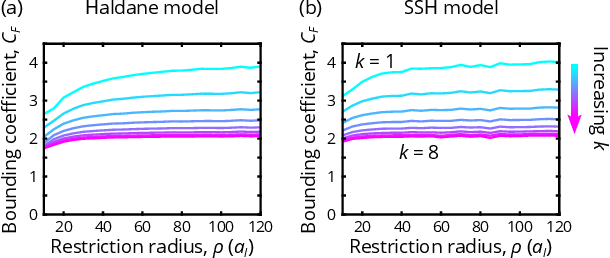}
    \caption{Estimation of $\rho \Vert [F_\rho(X),H] \Vert / \Vert [X,H] \Vert$ for the class of functions $\hat{\phi}_k(x)$ as a function of the restriction radius $\rho$ and choice of $k$ in $\hat{\phi}_k(x)$ for the Haldane lattice (left) and the SSH lattice (right). The Haldane lattice uses the parameters $t_c = t/2$, $\phi = \pm \pi/2$, and $M = 0$, while the SSH lattice uses $t_2 = 0.2t$. The value of $k$ increases from $1$ to $8$ in steps of $1$ as cyan turns to magenta.}
    \label{fig:numerCF}
\end{figure*}

\vspace{.2cm}

However, to get an improvement on $C_F=3$, one needs to change the class the functions $\phi$. The above indicates that one should roughly expect $C_F\approx 2\,(C_{\phi})^{-1}\,\phi(\tfrac{1}{2})$ where $\phi$ is symmetric around $\frac{1}{2}$ and increasing on $[0,\frac{1}{2}]$, but the choice of $\phi$ has to be such that it is small and sufficiently smooth at the boundaries $0$ and $1$ in order to assure that the boundary terms do not dominate (at least a linear decay is needed for this). Motivated by this insight, let consider the family of functions
$$
\hat{\phi}_k(x) \;=\; e^{-\frac{2^{-k}}{x}}\,e^{-\frac{2^{-k}}{1-x}} \;.
$$
They are exponentially small at the boundaries, so that the boundary errors are expected to be very small. On the other hand, for all $x\in(0,1)$, one has $\lim_{k\to\infty} \hat{\phi}_k(x)=1$. In particular, $C_{\hat{\phi}_k}\uparrow 1$ as $k\to\infty$. Therefore, one concludes that $C_F\downarrow 2$ as $k\to\infty$. In lieu of a generally applicable analytical technique to derive such an improved estimate, let us turn to numerical tests. This is given in Fig.~\ref{fig:numerCF} and clearly supports that $F$ can be chosen such that $C_F\approx 2$.

%%%%%%%%%%%%%%%%%%%%%%%%%%%%%%%%%%%%%%%%%%%%%
\section{Two-parameter analytic perturbation theory}
\label{app-TwoPar}

Let $(x,s)\in B_r(0,0)\subset\RM^2\mapsto L(x,s)$ be a real analytic family of selfadjoint operators on a Hilbert space. Suppose that the $0$ is not in the essential spectrum of $L(0,0)$. Then the same holds for the whole family, provided that the radius $r$ of the ball $B_r(0,0)$ is chosen sufficiently small. Furthermore, suppose that $0$ is a simple eigenvalue of $L(0,0)$. One can then analyze its perturbation theory by standard resolvent techniques developed in Kato's book \cite{Kat}, even though this book only deals with a single parameter. Let us collect the main outcomes for the eigenvalue $(x,s)\in B_r(0,0)\mapsto \mu(x,s)$ and its spectral projection $(x,s)\in B_r(0,0)\mapsto P(x,s)$. Both $\mu$ and $P$ are real analytic in both parameters and therefore have convergent series expansions:
$$
\mu(x,s)
\;=\;
\sum_{i,j\geq 0}\mu_{i,j}x^is^j
\;,
\qquad
P(x,s)
\;=\;
\sum_{i,j\geq 0}P_{i,j}x^is^j
\;,
$$
where $\mu_{i,j}\in\RM$, satisfying by assumption $\mu_{0,0}=0$, and $P_{i,j}$ are bounded selfadjoint operators with further properties, assuring that $P(x,s)$ is a projection (selfadjoint idempotent). In particular, this means that for $r$ sufficiently small the projection $P(x,s)$ remains one-dimensional. The algebraic relations on the selfadjoint coefficient operators $P_{i,j}$ follow from $P(x,s)^2=P(x,s)$, by comparing the terms on both sides. Let us focus mainly on the low order terms
$$
\mu(x,s)
\;=\;
\mu_{1,0}x+\mu_{0,1}s+\mu_{1,1}xs+\Oo(x^2,s^2)
\;,
\qquad
P(x,s)
\;=\;
P_{0,0}+P_{1,0}x+P_{0,1}s+\Oo(x^2,s^2,xs)
\;,
$$
where $\mu_{0,0}=0$ was used, implying that $L(0,0)P_{0,0}=0$. Then $P_{0,0}$ is a projection and
$$
(P_{0,0}+P_{1,0}x+P_{0,1}s)^2
\;=\;
P_{0,0}+P_{1,0}x+P_{0,1}s
+
\Oo(x^2,s^2,xs)
$$
shows
$$
P_{0,0}P_{1,0}(\one-P_{0,0})\;=\;0\;=\;
(\one-P_{0,0})P_{1,0}P_{0,0}
\;,
\qquad
P_{0,0}P_{0,1}(\one-P_{0,0})\;=\;0\;=\;
(\one-P_{0,0})P_{0,1}P_{0,0}
\;,
$$
namely $P_{0,1}$ and $P_{1,0}$ are both off-diagonal w.r.t.\ $P_{0,0}$. To prove the above statements and also to obtain explicit   formulas given below, one uses that $P(x,s)$ is given by the Riesz formula
$$
P(x,s)
\;=\;
\oint_\Gamma\frac{dz}{2\pi\imath}\,(z\one -L(x,s))^{-1}
\;,
$$
where $\Gamma$ is a positively oriented circle in the resolvent set of $L(x,s)$ encircling merely the eigenvalue $\mu(x,s)$. Now one can simply iteratively apply the resolvent identity to compute the coefficients $P_{i,j}$. For this purpose, suppose for sake of concreteness that
$$
L(x,s)\;=\;L+x V +s W
\;,
$$
where $V$ and $W$ are two selfadjoint operators that are relatively bounded w.r.t.\ $L$, namely $V(L-E\,\one)^{-1}$ and $W(L-E\,\one)^{-1}$ are bounded operators. Then
\begin{align*}
&
\!\!\! 
(z\one -L(x,s))^{-1}
\\
&
\;=\;
(z\one -L(0,s))^{-1}+(z\one -L(0,s))^{-1}xV(z\one -L(x,s))^{-1}
\\
&
\;=\;
(z\one -L(0,s))^{-1}+(z\one -L(0,s))^{-1}xV(z\one -L(0,s))^{-1}+\Oo(x^2)
\\
&
\;=\;
(z\one -L(0,s))^{-1}+(z\one -L)^{-1}xV(z\one -L)^{-1}+\Oo(x^2,xs)
\\
&
\;=\;
(z\one -L)^{-1}+(z\one -L)^{-1}s W(z\one -L)^{-1}+(z\one -L)^{-1}xV(z\one -L)^{-1}+\Oo(x^2,xs,s^2)
\;.
\end{align*}
Hence
$$
%P_{0,0} \;=\; \oint_\Gamma\frac{dz}{2\pi\imath}\,(z\one -L)^{-1} \;,
%\quad
P_{1,0}
\;=\;
\oint_\Gamma\frac{dz}{2\pi\imath}\,(z\one -L)^{-1}V(z\one -L)^{-1}
\;,
\quad
P_{0,1}
\;=\;
\oint_\Gamma\frac{dz}{2\pi\imath}\,(z\one -L)^{-1}W(z\one -L)^{-1}
\;.
$$
With this given, the eigenvalue equation
$$
L(x,s)P(x,s)
\;=\;
\mu(x,s)P(x,s)
\;
$$
expanded in both $x$ and $s$ gives, up to orders $\Oo(x^2,s^2)$,
\begin{align*}
&
x LP_{1,0}+s LP_{0,1}+xs LP_{1,1}
+xVP_{0,0}+s WP_{0,0}+xs VP_{0,1}+xs WP_{1,0}
\\
&
\;\;=\;
x\mu_{1,0}P_{0,0}+s\mu_{0,1}P_{0,0}+xs\mu_{1,1}P_{0,0}
+xs\mu_{1,0}P_{0,1}+xs\mu_{0,1}P_{1,0}
\,+\,\Oo(x^2,s^2)
\;.
\end{align*}
From the coefficients of $x$, one gets
\begin{equation}
\label{eq-FirstOrder}
LP_{1,0}+VP_{0,0}
\;=\;
\mu_{1,0}P_{0,0}
\;.
\end{equation}
Multiplying with $P_{0,0}$ and using $P_{0,0}L=(LP_{0,0})^*=0$, one gets 
$$
P_{0,0}VP_{0,0}\;=\;
\mu_{1,0}P_{0,0}
\;.
$$
If $P_{0,0}=|\phi\rangle\langle\phi|$ for a normalized vector $\phi$, this implies the well-known lowest-oder perturbation formula
$$
\mu_{1,0}
\;=\;
\langle \phi|V|\phi\rangle
\;.
$$
With this at hand, one can again use \eqref{eq-FirstOrder} to obtain
$$
LP_{1,0}
\;=\;
(\mu_{1,0}-V)P_{0,0}
\;=\;
-(\one-P_{0,0})VP_{0,0}
\;.
$$
so that one also obtains an alternative expression for $P_{1,0}$:
$$
P_{1,0}
\;=\;
-L^{-1}(\one-P_{0,0})VP_{0,0}
\;=\;
-(\one-P_{0,0})L^{-1}(\one-P_{0,0})VP_{0,0}
\;.
$$
Note that the inverse $L^{-1}$ in this expression indeed exists on the image of $\one-P_{0,0}$ and that it commutes with the projection $\one-P_{0,0}$, showing the second formula. Similarly
$$
\mu_{0,1}
\;=\;
\langle \phi|W|\phi\rangle
\;,
\qquad
P_{0,1}
\;=\;
-(\one-P_{0,0})L^{-1}(\one-P_{0,0})WP_{0,0}
\;.
$$
Note that this does not involve yet the first order terms in $P(x,s)$. For the computation of $\mu_{1,1}$, one does need to use them though, because
$$
LP_{1,1}
+VP_{0,1}+ WP_{1,0}
\;=\;
\mu_{1,1}P_{0,0}
+\mu_{1,0}P_{0,1}+\mu_{0,1}P_{1,0}
\;.
$$
Multiplying by the projection $P_{0,0}$ from the left and right and using the fact that $P_{0,1}$ and $P_{1,0}$ are off-diagonal as well as again $P_{0,0}L=0$, one gets
$$
P_{0,0}VP_{0,1}P_{0,0}\,+\, P_{0,0}WP_{1,0}P_{0,0}
\;=\;
\mu_{1,1}P_{0,0}
\;.
$$
Using again the normalized vector $\phi$ spanning $P_{0,0}$, one gets
\begin{align}
\mu_{1,1}
&
\;=\;
\langle \phi|VP_{0,1}|\phi\rangle
+
\langle \phi|WP_{1,0}|\phi\rangle
\nonumber
\\
&
\;=\;
-\langle \phi|V(\one-P_{0,0})L^{-1}(\one-P_{0,0})W|\phi\rangle
-
\langle \phi|W(\one-P_{0,0})L^{-1}(\one-P_{0,0})V|\phi\rangle
\nonumber
\\
&
\;=\;
-\,2\,\Re e\big(\langle \phi|V(\one-P_{0,0})L^{-1}(\one-P_{0,0})W|\phi\rangle\big)
\;.
\label{eq-SecondOrderPert}
\end{align}

\vspace{.3cm}

\noindent {\bf Acknowledgements:}  A.C.\ acknowledges support from the Laboratory Directed Research and Development program at Sandia National Laboratories. This work was performed in part at the Center for Integrated Nanotechnologies, an Office of Science User Facility operated for the U.S. Department of Energy (DOE) Office of Science. Sandia National Laboratories is a multimission laboratory managed and operated by National Technology \& Engineering Solutions of Sandia, LLC, a wholly owned subsidiary of Honeywell International, Inc., for the U.S. DOE's National Nuclear Security Administration under Contract No. DE-NA-0003525. The views expressed in the article do not necessarily represent the views of the U.S. DOE or the United States Government. The work of H.~S.-B. was supported by the DFG grant SCHU 1358/8-1.

%%%%%%%%%%%%%%%%%%%%%%%%%%%%%%%%%%%%%%%%%%%%%

\end{document}